# Peptides as versatile scaffolds for quantum computing

Lorena E. Rosaleny,*,† Alicia Forment-Aliaga,*,† Helena Prima-García,*,† Ramón Torres Cavanillas,† José J. Baldoví,‡ Violetta Gołębiewska,† Karolina Wlazło,† Garin Escorcia-Ariza,† Luis Escalera-Moreno,† Sergio Tatay,† Carolina García-Llácer,† Miguel Clemente-León,† Salvador Cardona-Serra,† Patricia Casino,¶ Luis Martínez-Gil,¶ Alejandro Gaita-Ariño,*,† and Eugenio Coronado†

†Inst. de Ciencia Molecular, U. de València, Cat. José Beltrán 2, 46980 Paterna, Spain
‡Nano-Bio Spectroscopy Group and ETSF, U. del País Vasco CFM CSIC-UPV/EHU-MPC & DIPC, Av. Tolosa 72, 20018 San Sebastián, Spain
¶Dept. de Bioquímica y Biología Molecular, U. de València, Dr. Moliner 50, 46100 Burjassot, Spain

E-mail: lorena.rosaleny@uv.es; alicia.forment@uv.es; helena.prima@uv.es; alejandro.gaita@uv.es

## Abstract

In this work we showcase the potential of peptides as versatile scaffolds for quantum computing and molecular spintronics. In particular, we focus on lanthanide-binding tags, which were originally developed in the field of biotechnology for the study of protein structure and dynamics. Firstly, we demonstrate quantum coherent oscillations in a Neodymium and Gadolinium peptidic qubits. Then, employing bacterial biosynthesis, we investigate the possibility of increasing the number of qubits in the same molecular system, with the case studies being a double spin qubit with two distinct coordination environments, and an asymmetric chain of 9 spin qubits with a spin-spin separation of about 2 nm and in an arbitrarily chosen sequence of coordination environments. Finally, we take advantage of biochemical modification for the preparation of paramagnetic, chiral, Self-Assembled Monolayers (SAMs) on Au(111). Our experimental and theoretical characterization shows that this is a promising structure for spintronic applications, and in particular to improve on two state-of-the-art approaches to molecular spin qubits. We conclude with an overview of the challenges and new opportunities opened by this emerging field.

## Introduction

In a world devoted to the processing of information, at an ever faster pace and with ever smaller devices, the effervescent field of spintronics strives to reach these targets by going beyong the charge degree of freedom of the electrons and profiting also from their spin degree of freedom.[1] Molecular spintronics adds to the general possibilities of conventional spintronics the intrinsic properties of molecules both for the chemical design of modular, functional molecular units including molecular spin qubits[2] and for the organisation of said molecules in layers which then act as barriers, spin filters or interface modifiers.[3]

In a straightforward application of molecular nanospintronics to quantum computing, it has been shown that single-molecule spin transistors can be used to measure and control individual nuclear spin qubits.[2,4] These experiments determine the state of nuclear spin qubits by measuring the change in the electric current passing through a molecule when its electronic spin is flipped by a continuously varying magnetic field. Currently this proposal works only with a single nuclear spin, and indeed a major limitation in potential quantum technologies is scalability. The role of molecules in this field is to offer 'smart' elementary building blocks that already include a minimal functionality, to be then wired up by a physical approach.[5,6] This means designing more complicated molecular structures as building blocks to construct complex molecular spintronic devices. Choosing biomolecules in this context would open the door to multiscale complexity, permitting the use of techniques developed in molecular biology, including recombinant protein technology[7] or CRISPR-CAS9.[8] These techniques enable obtaining relatively inexpensive, on-demand peptide sequences and modifications of existing proteins in a standard way, facilitating tailored molecular modification and multiscale organization. However, one needs to note that the use of spin-labelled proteins or polypeptides to implement spin-coherent phenomena, in artificial processes that imitate Quantum Biology,[9–11] can only have potentially far-reaching consequences in nanoscience and nanotechnology if these magnetic molecules can be shown to present spin coherence and be structurally stable in different environments. This would put metallopeptides in direct competition with purely chemical strategies previously reported for the organisation of spin qubits,[12,13] and in an interesting position to play a supporting role for the mixed chemical-physical approach for scaling of qubits, such as the coupling of individual multi-qubit molecules with superconducting transmission lines.[5,6,14] As a very recent advance in this direction, simple radical-labelled peptides have already been used to form a small molecular spin quantum network which can in turn be controlled by a nitrogen-vacancy center in diamond,[15] where however the radical presents some stability problems and is currently limited to dimers. The preparation of biomolecules with a larger number number of stable spin qubits in controlled positions would be the next natural step of this approach. From a wider perspective, the successful performance of spintronic devices relies largely on the control of spin-selective transport, and room temperature spin dependent transport has already been demonstrated on SAMs of DNA and of other more simple chiral molecules on gold electrodes, an effect that has been dubbed as Chirality-Induced Spin Selectivity (CISS).[16]

In this work we aim to advance towards the preparation of coherent, scalable, single-molecule spin transistors. This goal comprises three main phases: (1) characterization of a single, coherent, peptidic spin qubit, (2) demonstration of a certain control over the coordination environment and advancement towards higher nuclearities, and thus to a much higher number of qubits in a single biomolecular entity and (3) organization, for example on a surface, to allow electrical addressing. The manuscript will commence by describing Lanthanide Binding Tags,[24] which are the particular metallopeptides on which we will focus as case study, and demonstrating that they can be considered autonomously folding domains,[25] meaning their secondary structure does not critically depend on their environment. We will proceed to show their potential as spin qubits, in terms of quantum coherence. Moreover, we will comment on the procedure of property optimization via peptidic screening. We will then demonstrate the design of more complex biomolecular prototypes for quantum gates via bacterial biosynthesis. This is done by preparing a double assymetric coordination environment in a single peptide. The approach can be trivially extended to a larger nuclearity: the design of assymetric peptides with up to nonanuclear binding abilities will be showcased as example. Finally, we will illustrate how a biochemically trivial modification of the peptide sequence al-lows the formation of a paramagnetic, chiral, single-molecule-thin layer with potential interest in molecular spintronics. We will accompany the preparation and experimental characterization of this SAM with theoretical calculation on its conducting pathways, its potential for spin filtering and the predicted magnetic field steps that the paramagnetic ion creates along the conducting pathway.

## Results and discussion

### Biomolecular spin qubit

Employing biomolecules as hardware for spintronics is a general strategy that can conceivably be implemented in many different ways: any metalloprotein is in principle a candidate, as are radical-tagged peptides.[15] Equivalently, DNA origami[17] could be used for this same goal. As a proof of concept, in this work we need to focus on a particular case, both in terms of the spin-carrying moiety and in terms of the biomolecule.

For the spin moiety, here we chose lanthanide ions, which have demonstrated interesting results lately, such as magnetic hysteresis at 60 K,[18] implementation of Atomic Clock Transitions,[19] integration of a three-qubit quantum processor in a single ion,[5] and, crucially, spintronic readout of nuclear spin qubits at the single-molecular level.[2] In general, lanthanide ions are interesting candidates in the context of quantum technologies.[20] In particular, in this work we tested trivalent lanthanide ions Ln = Nd, Gd, Tb, Dy, Ho, Er, Tm, Yb.

On the side of the biomolecule, we chose to focus on the short peptide sequences (15-17 aminoacids) known as Lanthanide Binding Tags (see Fig. 1).[21] These were originally optimized by Imperali et al. via the exploration of a series of peptide libraries, starting from calcium-binding motifs of proteins, with the combined goals of fine-tuning the selectivity of binding to lanthanide ions and presenting photoluminescence.[23] As any short peptide, Lanthanide Binding Tags can easily be genetically encoded, thus they are used as luminescent spin tags in controlled protein positions.[24] In a recent example, a double tagging was used to perform non-disruptive in-cell structural measurements via Double Electron-Electron Resonance (DEER).[26] Herein we postulate that this can be extended to other applications requiring quantum coherence.

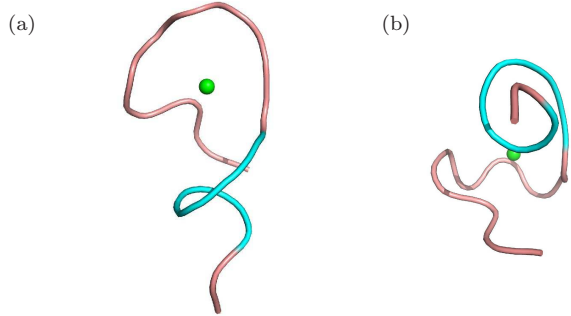

Figure 1: Structure of the lanthanide-binding region. (a) Lateral and (b) upper views from the trace of the secondary structure of **TbLBT** (PDB code: 1TJB). Processing of the structure with the STRIDE software determines a turn at the N-terminus and a short $3_{10}$-helix at the C-terminus (marked in cyan).[27]

The most standard Lanthanide Binding Tag is YIDTNNDGWYEGDELLA (from now on: **LBT**); moreover several peptides that also bind lanthanides have sequences that are closely related to it (see Table 1), thus we employed it as a typical sequence to test for quantum coherence. In this work, all **LnLBT** samples were prepared from commercially acquired **LBT** and spectroscopically characterized as described in Methods and Supporting Information section S1.

The relaxation of the spin implies a loss of quantum information and thus is undesirable for our purposes. It is usually characterized by two parameters: spin-lattice relaxation time $T_1$ and spin-spin relaxation time $T_2$. Both are detected in spin echo experiments, using different sequences of pulsed EPR. $T_1$ is usually longer and related with vibrational dissipative processes, hence it tends to have a large temperature dependence. $T_2$ is usually shorter and is related with more subtle processes governed by the so-called spin bath, which is the collection of nuclear and electronic spins that constitute the magnetic environment of the spin we are characterizing in our EPR experiment.

Table 1: Previously reported Lanthanide Binding Tags X-ray resolved crystal structures (except for 2LR2 which was determined by NMR). Note that dozens of other variations of the same sequence have been characterized as selectively binding to lanthanide ions,[23] but most have not been crystallized.

| protein | Ln | # indep. sites | LBT aminoacid sequence | resolution (Å) | PDB ID[ref] |
|---|---|---|---|---|---|
| **LBT** | Tb | 2 | YIDTNNDGWYEGDELLA | 2.0 | 1TJB[28] |
| dLBT-Ubiquitin | Tb | 2 | GPGYIDTNNDGWIEGDEL––YIDTNNDGWIEGDELLA | 2.6 | 2OJR[29] |
| Interleukin-1$\beta$-(S1-LBT) | Tb | 1 | GYIDTNNDGWIEGDELY | 2.1 | 3LTQ[30] |
| Interleukin-1$\beta$-(L3-LBT) | none | 1 | GYIDTNNDGWIEGDELY | 1.7 | 3POK[30] |
| xq-dSE3-ubiquitin | Gd | 4 | YIDTDNDGSDGDEL––YIDTDNDGSIDGDELLA | 2.4 | 3VDZ[31] |
| Z-L2LBT | none | 1 | SYIDTNNDGAYEGDELSG | n.a. | 2LR2[32] |

Peptides have a high number of vibrational degrees of freedom and a high number of nuclear spins (from H atoms), so one would naïvely expect short relaxation times compared to regular coordination compounds, however a previous theoretical study considering the $^1$H nuclear spin bath as decoherence sorce estimated an upper limit of the decoherence times above tens of $\mu$s in several members of the **LnLBT** series,[33] and indeed previous experimental results indicate coherence in these kind of complexes.[34]

In the preliminary continuous-wave EPR exploration, we only obtained signal for **GdLBT** and **NdLBT**. For these two derivatives we also found a signal in the pulsed experiment, see Fig. 2 and Supporting Information section S2. The relaxation times were found to be at least comparable to those typical for Ln complexes,[5] with $T_1 \simeq 10$ $\mu$s, $T_2 \simeq 2$ $\mu$s in the case of **GdLBT** (for further details, see Supporting Information section S2). Note that this is also consistent with recent DEER measurements in Gd-peptide tagged proteins[34] were $T_2 \simeq 2$ $\mu$s had also been measured. In the case of **NdLBT**, the relaxation times are somewhat shorter, $T_1 \simeq 3$ $\mu$s and $T_2 \simeq 1$ $\mu$s, and yet it is remarkable that this is, to the best of our knowledge, the first reported Nd-based molecular spin qubit. The contribution of the proton spin bath was estimated to be moderate, providing a ceiling for the coherence time in the order of $T_2 < 200\mu$s, meaning it is not the limiting factor in this case, see Supporting Information section S2.2.

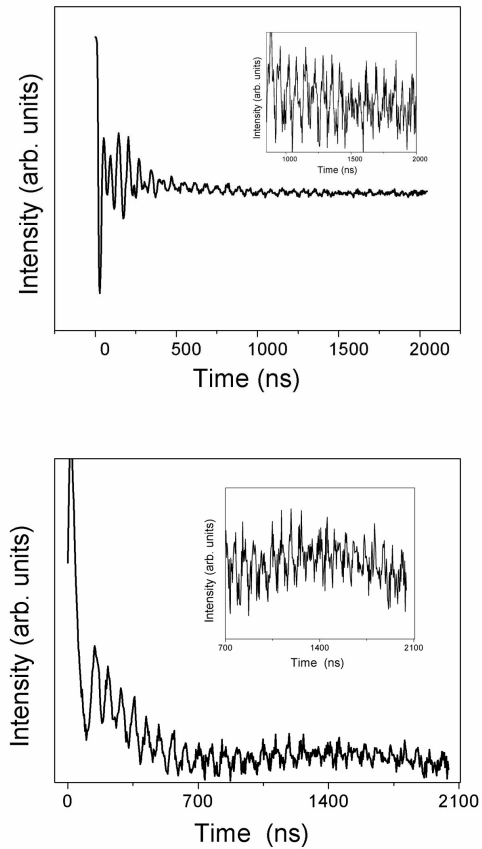

Figure 2: Coherent Rabi oscillations at the Hartmann-Hahn condition in **GdLBT** (up) and **NdLBT** (down).

One needs to note here that sufficiently long $T_1$ and $T_2$ times are a necessity, but not a guarantee, since these parameters only quantify the survival of quantum information that is not actually being manipulated, which is not a practical setup. Rabi oscillations are much more interesting: they are the coherent back-and-for oscillations between two quantum states, like clock cycles. We may picture Rabi oscillations as consecutive "NOT" logical operators that invert the value of the qubit over and over. In very noisy systems, such as most solid-state systems, quantum coherence is lost in such a short timescale that Rabi oscillations are impossible: detecting Rabi oscillations for at least a few cycles is considered a *sine qua non* condition for the use of a particular hardware for quantum applications. We studied Rabi oscillations at the Hartmann-Hahn conditions for best performance,[22] and characterized approximately 20 coherent oscillations, both for **GdLBT** and for **NdLBT**, meaning a maximum of 20 quantum operations can be performed before losing all vestiges of coherence.

Considering both $T_2$ values and the number of Rabi oscillations, it is clear that they are still too low to be practical. At the same time, these numbers establish that peptidic qubits are already in the same range of coherence as the majority of non-optimized complexes from coordination chemistry. Of course, the main advantage of biomolecular qubits is not high coherence but to offer a support for organisation at the nanoscale. Nevertheless, a further benefit of peptides is systematic property optimization: whether one is interested in quantum coherence or in any other specific spintronic effect, the rational procedure to improve the desired property would be to perform a combinatory screening of a peptide library, i.e. to make combinations of substitutions of individual aminoacids in the sequence to obtain progressively better properties. This kind of procedure is precisely what originated the original **LBT** sequence, as an optimization of lanthanide affinity starting from $Ca^{2+}$-binding motifs of proteins.[23] In that case the dissociation constant $K_D$ was reduced by almost three orders of magnitude, from $K_D = 9$ $\mu$M to $K_D = 19$ nM, in a procedure that involved exploring seven consecutive peptide libraries and resulted in the substitution of 6 aminoacids in the sequence and the addition of two further ones. It is reasonable to expect that a similar procedure will result in a notable enhancement of the properties of our interest; the procedure could be guided by combining recent theoretical advancements that determine the key vibrations that couple to spins[35] with inexpensive effective electrostatic calculations of the crystal field.[36] For a lengthier discussion see Supporting Information section S3.

Dc magnetometry was employed to gain some insight into the nature of the ligand field in these systems (see Supporting Information section S4). The **GdLBT** magnetic data (not shown) showed a perfect agreement with a Brillouin function, since its crystal field splitting typically corresponds to temperatures below 2 K. Additionally, **TbLBT**, **DyLBT** were studied because of their relevance in molecular magnetism, since they present a large total magnetic moment $J$. Like **NdLBT**, these derivatives present the typical decreasing of $\chi T$ when lowering the temperature due to the depopulation of the Stark levels that result from the ligand field. A theoretical fit of the magnetic properties resulted in estimates for crystal-field parameters and ground-state wavefunctions (see details in the Supporting Information section S5).

## Biomolecular prototypes for quantum gates

Having established the potential of single peptidic qubits, the next step is working towards multi-qubit biomolecules, or in other words biomolecular prototypes for quantum gates and minimal quantum logic. The key here will be to combine two or more instances of **LBT**, with differences in the coordination environment: we need to combine in a single molecule two sig-nificantly different coordination environments, effectively achieving a dissimetric dinuclear lanthanide complex, a known pathway to implement quantum gates.[38]

However, as a previous step towards this goal, we need to demonstrate autonomous folding of **LBT**: in general, the folding of a sequence of aminoacids depends on its environment so multiple copies of a given peptide can in principle fold in different ways. Using the TopMatch software for alignment and superposition between peptidic sequences, it is possible to support the claim that the 15 aminoacid sequence YIDTNNDGWYEGDEL is an autonomous folding region, since is able to self-organise in a predictable form in different crystal structures, whether it is in free form, or fused with an ubiquitin protein, Fig. 3(a) or inserted in a loop of interleukin-1-$\beta$, Fig. 3(b). The SHAPE software also confirms small geometric distances of the 15 $\alpha$ carbons of the peptidic sequence, when comparing the different crystallographically determined structures (see Supporting Information section S6).

Moreover, with the SHAPE software we also determined that the 8 oxygens atoms from the coordination sphere are, up to a certain point, also kept constant between different crystal structures, as long as the sequence of aminoacids is conserved (see Supporting Information section S6). In all cases the coordination environment consists in a combination of mono- and bi-dentated carboxylate/carboxamide residues (aspartate, asparagine or glutamate). As a side note, the coordination is completed with the participation of the oxygen of an amide bond, meaning that the contact of the lanthanide with the peptide chain does not only occur via the lateral groups but also directly; as we will see below this can be important for its spintronic potential since a current flowing through the peptide will be directly exposed to the magnetic field of the lanthanide ion.

Because of the robustness of the folding in these peptides, it has been shown that it is possible to substitute a carboxamide by a carboxylate (asparagine by aspartate) with no change in the structure (see Table 1). For our purposes, this means that it is possible to make controlled changes in the coordination environment, altering the distribution of the charges but maintaining the locations of the ligands. Thus, as a double-qubit peptidic prototype we employed YIDTDNDGWYEGDELYIDTNNDGWYEGDELLA (double assymetric LBT or in short: **daLBT**). This is a novel fusion of two known lanthanide binding subunits with the mentioned carboxamide-carboxylate modification in the coordination sphere. Thus, the residue marked as asparagine N3 in Figure 3 (c) is occupied by carboxylate in the first unit and carboxamide in the second unit, resulting in two different coordination environments. In this dinuclear system we expect a Ln-Ln distance of $20 \pm 1$Å (see Supporting Information section S7), in the order of the distance in latest experiments of qubit networks employing simpler peptides.[15]

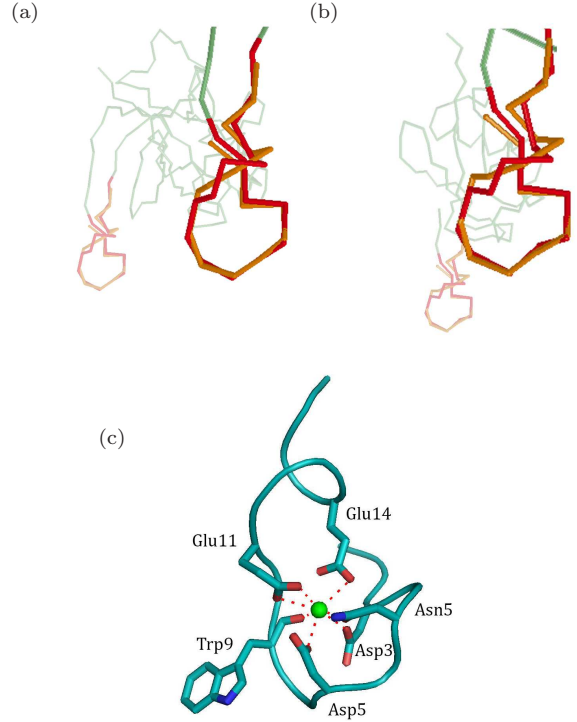

Figure 3: Alignment and superposition, calculated using TopMatch[39] of 1TJB (in orange) with similar sequences either fused with a ubiquitin (a) or inserted in the middle S-loop of an interleukin-1-$\beta$ (b), highlighting the red matching region in the otherwise green structure. Note the very similar folding of the region (highlighted, front) in completely different proteins (transparent, back). (c) $C_\alpha$ trace from peptide 1TJB with metal-coordinating residues represented as sticks, terbium as a green sphere, and red dashes as metal-O bonds.

For the preparation of this novel dimer we employed bacterial biosynthesis, or recombinant protein expression, a standard Molecular Biology technique to allow production of target proteins. It is an affordable alternative to chemical synthesis for long peptides/proteins (longer than a few decades of aminoacids). For our purposes, this technique will be essential when the production of longer peptides is required i.e. for peptides containing multiple Lanthanide Binding Tags. Typically, in recombinant protein expression, the DNA sequence coding the target protein is placed under the control of an inducible promotor, meaning that expression only occurs when a specific chemical compound is present in the culture media. This DNA sequence is then inserted into a commercial plasmid (an auto-replicative circle of DNA that propagates in bacteria). When the control compound is added to a culture of plasmid-containing bacteria, the protein is expressed. Standard peptidic tags, that are already in the plasmid, fused in N- or C-terminal position to the protein of interest, allow the purification of the target protein via Affinity Chromatography.

Bacterial biosynthesis, followed by purification by chromatography and analysis by electrophoresis resulted in the detection of a peptide with a molecular weight $M_w \simeq 30$ kDa, compatible with the predicted recombinant protein ($M_w(\text{GST}) = 26$ kDa + $M_w(\textbf{daLBT}) = 4$ kDa). In a posterior Western-Blot analysis with an anti-GST antibody, the putative GST-**daLBT** band produced luminiscence, further supporting the hypothesis that GST-**daLBT** has been obtained. Further details are given in Supporting Information section S7.

This method allows the extension of the synthesis for longer sequences, acommodating a higher number of spins. As an example, we illustrate the procedure to obtain polypeptides that range from mononuclear to nonanuclear in two diferent coordination environments. A DNA sequence coding for a nonanuclear complex was designed, arbitrarily choosing in a 010011101 sequence where 0 and 1 correspond to the carboxylate and the carbamide variations of **LBT** described above, respectively. Unique linkers were used, based on permutations of the aminoacid sequence GASAG. The election of G, A and S was made to facilitate a free arrangement of each site and thus not impede their independent folding (see Supporting Information section S9 for details including the full DNA sequence). The unique spacers allow for specific PCR amplification of peptides coordinating an arbitrary number of ions between 1 and 9. Amplified DNA could be then sub-cloned, fused to the GST and purified in *E. coli* as demonstrated above in the simple case of dimers.

Note that there are other possibilities to apply the peptidic approach to obtain a complex scaffold for the 3D organisation of molecular spintronics components. For reasons of space, several more sophisticated approaches, either purely based on biomolecules or combining biomolecules with inorganic objects known as polyoxometalates, are discussed as Supporting Information section S10.

## Self-assembled monolayers of biomolecular qubits

Having studied the potential of these robust peptidic motifs for coherent quantum manipulations and the possibility to control the nuclearity and thus the number of qubits, the next step is organizing them on a surface. At the same time, it is worth considering their possibilities from an ample spintronics perspective. Firstly, the intrinsic chirality of LBT units and the additional presence of a paramagnetic lanthanide working as a source of an internal magnetic field, make them interesting candidates for their use as potential spin filter layers in spintronic devices with potential CISS effect. Secondly, as suggested in previous works, the mere combination of a conducting magnetic molecule where the magnetic ion presents a nuclear spin and where the electronic spin states survive in the conduction regime could offer possibilities for the electrical reading of nuclear spin qubits.[2,4] Since these conditions might be met in the case of **TbLBT**, these ideas motivated us to study the possibility of organizing the **LnLBT** units as one molecule thick films on gold surfaces, that is, the formation of **LnLBT** SAMs.

The self-assembling of molecules on gold is a well known strategy driven by specific interactions between a functional group in the molecular unit with the metallic surface and additional weak interactions (van der Waals, hydrophobic...) among the skeleton of the molecules, giving rise to a well organized film. Over years, thiol group has been extensively used to anchor molecules to bare metals.[40] In the case of peptides, the inclusion of a thiol group can be achieved trivially, by adding a cysteine aminoacid to a known sequence. This is the reason why, for this part of the study, we employed YIDTNNDGWYEGDELC (short: **LBTC**), a novel modification of **LBT** where we substituted two aminoacids, that had been shown not to be critical for lanthanide binding, by a single cysteine. **LBTC** was prepared by Solid Phase Peptide Synthesis and acquired commercially, and subsequently spectroscopically characterized by us as described in Methods and Supporting Information section S1.

The standard SAM growth procedure consists on immersion of clean gold substrates in diluted solutions of the target molecule followed by a cleaning step with net solvent to remove physisorbed molecules. In order to avoid interference with the SAM formation, we employed a pH buffer with no sulfur atoms (see Supporting Information section S11). After the treatment, the topography of the functionalized substrate was studied by Atomic Force Microscopy (AFM), in order to rule out the presence of aggregates on the surface. After functionalization, the roughness remains almost unchanged suggesting a homogeneous coverage as can be seen in Fig. 4(a) where the topographic AFM image of a substrate functionalized with **TbLBTC** SAM is compared with a reference sample (incubated overnight in a buffered solution without **LBTC**). To confirm the presence of **LBTC** on the gold surface, Matrix-Assisted Laser Desorption/Ionization Time-Of-Flight Mass Spectrometry (MALDI-TOF) was carried out (see Supporting Information section S12). As can be seen in Fig. 4(b), a main peak at $m/z = 1630$ corresponding to a [DTNNDGWYEGDELC]$^+$ fragment can be clearly detected, (**LBTC** losing the first two N-terminal aminoacids). Moreover, the presence of a small peak at $m/z = 1929$ can be assigned to the whole **LBTC** unit with an additional Na$^+$ cation, proving the integrity of the molecule on the surface. Due to the weak coordination bonds existing between the **LBTC** and the Tb$^{3+}$ cation, it is not possible to detect the presence on the surface of this Lanthanide by MALDI-TOF analysis because Tb$^{3+}$ disentangles from the peptide during the ionization process. To demonstrate the presence of the Tb$^{3+}$ ion in the SAM, X-ray photoemission spectroscopy (XPS) was used. XPS spectra clearly show the presence of Tb3d peaks (1277 and 1242 eV, see Supporting Information section S13) denoting the presence of the lanthanide on the surface, besides peaks of N, C and S from **LBTC**. Finally, in order to evaluate the coverage reached with the formation of the SAM, a Quartz Crystal Microbalance was used (see Supporting Information section S14). This technique has been extensively applied to detect mass changes at sub-nanogram level.[41] As observed in Fig. 4, when **TbLBT** is anchored on the electrode, a frequency shift of 20 Hz is measured corresponding with a coverage range of 81-116% depending on the model used for a full coverage estimation. This is consistent with the topography results and points towards a good coverage for the SAM.

The combination of magnetism and helicity that is present in metal complexes of **LBTC** attached to Au is promising towards the observation of the CISS effect and/or magnetorresistance, and indeed peptidic design has recently been shown to be useful to optimize the CISS.[45] Moreover, as we proposed recently,[4] the spintronic readout of nuclear spin qubits at the single-molecular level should be feasible in many molecules, with the critical condition that the conduction path does not overlap with the magnetic orbitals. We therefore performed theoretical calculations to start to assess the potential of **LnLBTC** for quantum applications. As a previous step, we employed Density Functional Theory (DFT) calculations to confirm the conservation of the characteristic folding of the peptide chain upon attachment to Au(111) via the thiol group (geometric distance $S \leq 3$, see Supporting Information section S15). NEGF-DFT calculations have been performed on the sandwiched system obtaining a zero-bias transmission spectra with the nearest transmission peaks located at $E - E_F = -0.3$ eV (Fig. 4(d)), with the Local DOS plot for such state being represented at Fig. 4(e). This means either a moderate gate voltage or a strong bias will likely to be needed to guarantee conductance. The fact that the widening of the transmission peaks is rather moderate means a low hybridization between the biomolecule and the metallic electrode.

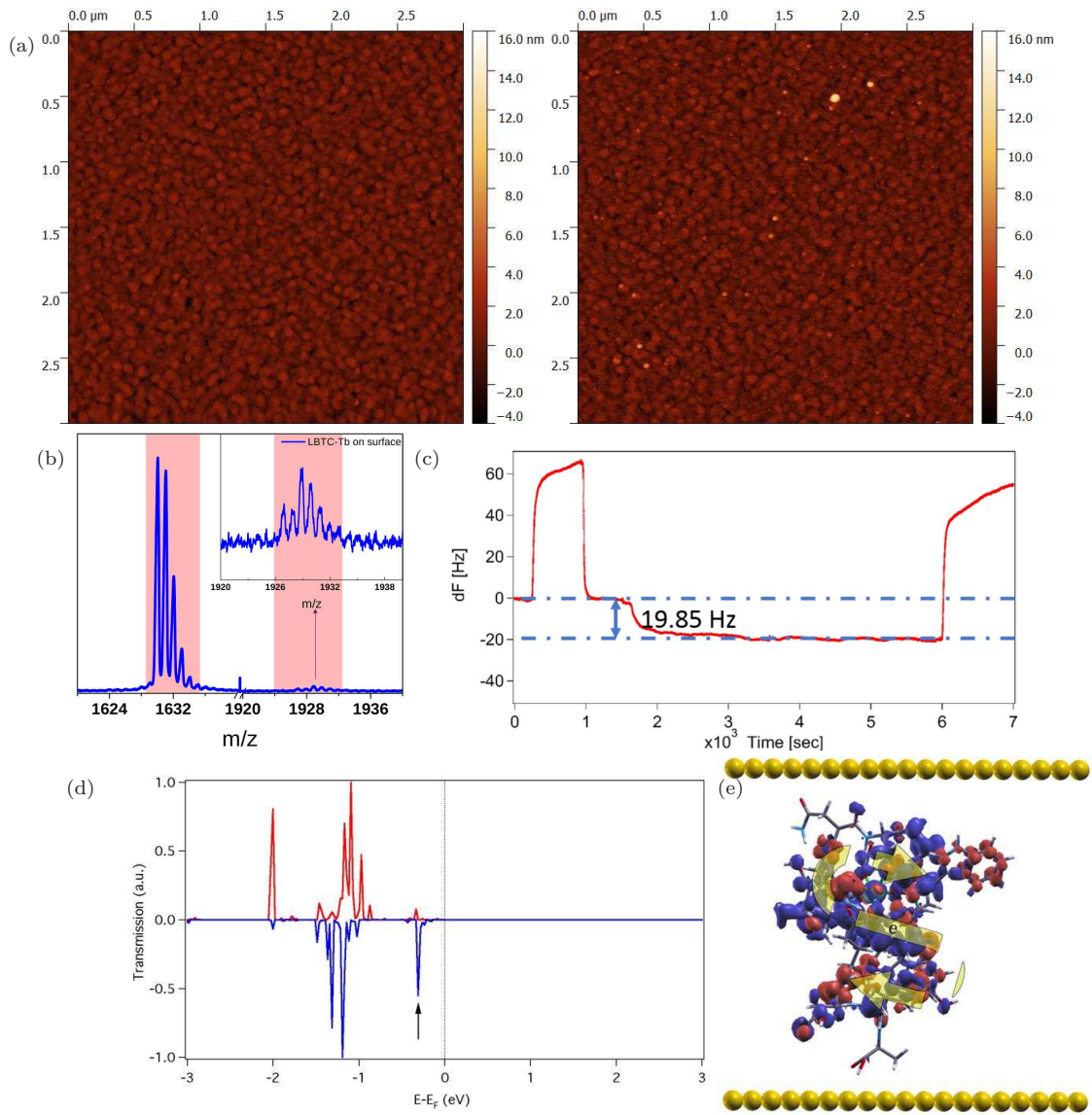

Figure 4: (a) AFM images of a gold surface incubated overnight with buffered solution (HEPES 10mM pH 7.0, NaCl 100mM, TbCl$_3$ 0.11 mM, TCEP 0.5mM) (left), or with buffered solution 0.1mM LBTC (right). (b) MALDI-TOF spectra measured for **TbLBTC** SAM: fragment of $M_w = 1630$, inset: complete sequence plus Na$^+$ ion, $M_w = 1929$. (c) Quartz Crystal Microbalance determination of the coverage via the shift in vibrational frequency $\Delta f$. (d) Calculated transmission spectra. A sharp spin-down conduction peak near the Fermi level is marked with an arrow as the main conducting peak. (e) Local Density of States of the main conducting peak. Most of the density is located in the $\pi$ orbitals of the peptide amide bonds.

Let us now consider two limit regimes characterized by the conduction via the $\pi$ clouds of the amide bond backbone, which are expected to be strongly involved in conduction.[42] In the first regime the channel is an extended, continuous SOMO that is bonded to the lanthanide ion via the amide coordination, with a typical range of exchange coupling $J = 1$–$25\,\mathrm{cm}^{-1}$.[43] In this regime and in the presence of CISS one should expect magnetorresistance at temperatures where the lanthanide polarization determines the polarization of the peptide SOMO. In the second regime, conduction takes place in discrete tunneling steps between amide $\pi$ bond (see Fig 4(e)). As reported previously for a carbon nanotube in presence of local fields,[44] this can be seen as a series of spin-polarizer+spin-analizer steps. We calculated the magnetic fields generated by the lanthanide ion at the different NCO amide $\pi$ bonds (see Supporting Information section S16). The results depend on the values and orientation of the Landé g factors, but assuming full spin polarization of the lanthanide, the field felt by the NCO group directly coordinated to it is of the order of 1 T, while for the rest of the chain the values are one or two orders of magnitude lower.

# Conclusions

Peptides known as Lanthanide Binding Tags were developed in the field of Biochemistry for the study of protein structure and dynamics. Herein we have expanded their applications and shown that they can be used as spin tags to open a promising playground in Molecular Spintronics, and in particular for improving current quantum computing schemes: both the one based on the electrical reading of nuclear spin qubits and the one based on the NV centers to manipulate small spin qubit networks. Among the advantages of employing metallopeptides as spin qubits are the ease of preparing new structures and the possibility to organise them in space, using techniques of Molecular Biology. Our first tests of organisation, namely complex dimeric/polymeric molecules with two different coordination sites via bateral biosynthesis, and preparing a Self Assembled Monolayer via biochemical modification were successful, if preliminary. A further key advantage of peptides as qubits is property optimization via the combinatorial screening of peptide libraries: in principle, it should be possible to minimize the coupling between the spin qubit and molecular vibrations by finding the sequence of aminoacids where low-energy vibrations do not alter the coordination environment of the magnetic metal. One needs to point out that peptides do present disadvantages: in particular crystallization is usually challenging in peptides, and indeed it was not achieved in this case, even after applying a wide array of crystallization conditions (see Supporting Information section S17). Nevertheless, this work proves that metallopeptides open different avenues for molecular spintronics or quantum computing that are closed to conventional coordination chemistry. Finally, since polyoxometalates have been found to bind regioselectively to positively-charged regions of different proteins,[46] one should consider the possibility of combining the advantages of polyoxometalates (see Supporting Information section S10),[5,19,22] such as their rigid structures and their absence of nuclear spins, two factors that favour high quantum coherence, with the organising power of proteins.

# Methods

## Biomolecular spin qubit

Peptides **LBT** and **LBTC** were purchased from Genscript. Peptides were synthesized by the stepwise Solid Phase Peptide Synthesis (SPPS) chemical method and quality-checked by MS and HPLC by the manufacturer.

$Ln^{3+}$ stock solutions were prepared from the $LnCl_3$ hydrate salts from Sigma-Aldrich (as 50 mM solutions in 1 mM HCl) and were diluted as needed.

To characterize **LBT** complexes we relied on UV-visible absorption and photoluminescence (see Supporting Information section S1). Absorption spectroscopy, where the extinction coefficient can be related with the concentration of tryptophan and tyrosine, was employed to confirm the purity of the peptide; the tryptophan-sensitized $Tb^{3+}$ photoluminescence was employed to confirm the binding to $Tb^{3+}$. Crystallization of **LBT** was attempted but proved challenging despite extensive efforts (see Supporting Information section S17).

For all spin studies we employed solutions with HEPES (4-(2-hydroxyethyl)-1-piperazineethanesulfonic acid) as pH buffer, a reducing agent and NaCl to preserve the integrity of the peptides. Since we employed cryogenic temperatures, we were forced to use either a frozen solution cryoprotected with glycerol, or liophilized powder.

To characterize the quantum spin dynamics of $Ln^{3+}$ ions encapsulated by **LBT** (**LnLBT**) we employed both continuous-wave and pulsed Electronic Paramagnetic Resonance (EPR) in frozen solution (see Supporting Information section S2 for details).

Dc magnetometry was performed on lyofilized samples of **GdLBT**, **TbLBT** and **DyLBT** at temperatures down to 2 K and magnetic fields up to 1 T (see Supporting Information section S5 for details).

For the theoretical analysis of the experimental $\chi T$ vs T data from the Tb and Dy derivatives, static magnetic properties were modeled employing a phenomenological approach using the full Hamiltonian in the CONDON package[37]

## Biomolecular prototypes for quantum gates

For the structure alignment we employed the web interface of TopMatch with the intermediate precision ("Match"). We compared the pairs (1TJB, 2OJR) and (1TJB, 3LTQ).

For the bacterial biosynthesis we employed a pGEX-2T plasmid with a GST tag in a culture of *E. coli* (see Supporting Information section S8 for details). After cell harvesting by centrifugation, sonication for cell lysis and another centrifugation step to get rid of cellular debris and other cellular structures, the supernatant underwent Affinity Chromatography using a Glutathion Sepharose 4B resin to retain the GST-tagged **daLBT** peptide. The purified peptide was analysed by Polyacrylamide Gel Electroforesis (SDS-PAGE).

## Self-assembled monolayers of biomolecular qubits

We used a standard solution with HEPES (4-(2-hydroxyethyl)-1-piperazineethanesulfonic acid) pH 7.0 as buffer, but substituted 2-mercaptoethanol by tris(2-carboxyethyl) phosphine (TCEP). To guarantee that **LBTC** coordinates the $Ln^{3+}$ cation when **LBTC**, TCEP and $Ln^{3+}$ ions are in the same solution, we worked always with an excess of the lanthanide and we verified by UV spectroscopy the presence of the expected luminescence emission at 544 nm induced by the **LnLBTC** complex when it is excited at 280 nm (see Supporting Information section S1). A simple immersion method was used to form a self-assembled monolayer: the gold substrate was cleaned and activated by piranha treatment, then the substrate was incubated in a buffered **TbLBTC** aqueous solution (HEPES 10mM pH 7.0 , NaCl 100mM, $TbCl_3$ 0.11 mM, TCEP 0.5mM, **LBTC** 0.1mM) overnight, and finally in order to remove non-chemisorbed material that could remain on the surface, the sample was introduced in water for two hours and copiously rinsed (details in Supporting Information section S11). After the treatment, the topology of the functionalized substrate was studied by Atomic Force Microscopy (AFM). To determine the presence of **TbLBTC** on the gold surface Matrix-assisted laser desorption/ionization time of flight (MALDI-TOF) spectrometry (Supporting Information section S12) and X-ray photoelectron (XPS spectroscopy) (Supporting Information section S13) were performed. To quantify the coverage we employed a Quartz Microbalance; in a typical experiment, a freshly cleaned Au covered piezoelectric quartz crystal was immersed in a liquid cell where a flow of different liquids at controlled rates was pumped while the oscillation frequency of the crystal was continuously monitored (see Supporting Information section S14). Several cycles of pure water and buffered aqueous solution were performed until the frequency was stabilized. Then right after a buffered solution step, **TbLBTC** was introduced during 60 minutes to ensure a constant frequency value. Finally, a flow of the buffered aqueous solution was injected to remove physisorbed **TbLBTC** on the crystal. Sauerbreys equation was used to estimate the relation between resonant frequency and mass loading of electrodes (see Supporting Information section S14).

# Acknowledgements

The present work has been funded by the EU (ERC Consolidator Grant DECRESIM and COST 15128 Molecular Spintronics Project), the Spanish MINECO (grant MAT2014-56143 and grant CTQ2014-52758 cofinanced by FEDER and Excellence Unit María de Maeztu MDM-2015-0538), and the Generalitat Valenciana (Prometeo Program of Excellence). A.G.A. acknowledges funding by the MINECO (Ramón y Cajal Program), L.E.M thanks the Generalitat Valenciana for an VALI+D predoctoral grant, L.M-G is currently funded by the MINECO (Juan de la Cierva Program). We thank Samuel Mañas-Valero and José V. Usagre for their assistance in the preparation of the EPR samples, as well as Fernando Coloma (SSTTI Universidad de Alicante) for the XPS measurements. The MALDI-TOF analysis was carried out in the SCSIE Universitat de Valncia Proteomics Unit, a member of ISCIII Proteo-RedProteomics Platform by O. Antnez and L. Valero.

# Associated content

## Supporting Information

The Supporting Information is available free of charge on the ACS Publications website at DOI: XXXXXXXX.XXX Additional information on several experimental and theoretical details, and also on future perspectives are included in the Supporting Information document. The document includes a spectroscopic characterization of the peptides used for EPR and SAMs experiments, additional CW EPR and pulsed EPR experimental characterization, magnetic measures (dc) and their theoretical fit (CONDON), considerations related to folding patterns (SHAPE) and 3D orientation of lanthanide centers in LBTs and double LBTs, details on the design of a nonanuclear LBT-based spin qubit, prospects of using biomolecules to organize Rare Earths, details on the bacterial biosynthesis of LBT peptides, the preparation method and characterization of LBTC SAMs (MALDI-TOF, XPS, QCM), transport calculations (LBTC) between two gold electrodes, theoretical estimates on the internal magnetic field created by the lanthanide on the peptide, and a summary of the crystallization efforts that have been done.

# Author contributions

LER, MCL, CGL characterized the LBT. HPG did the EPR experiments. AFA, RTC, LER, VG, KW, GAE, ST prepared and characterized the SAMs. SCS performed SMEAGOL transport calculations. JJB performed CONDON calculations on magnetic properties. LEM performed EasySpin fits and used SIMPRE to estimate magnetic fields and relaxation times. LER, PC, LMG performed the bacterial biosynthesis. AGA, LER, AFA and HPG conceived and supervised the work. AGA and EC prepared the manuscript with input from all authors. All authors contributed their reports to the Supporting Information, which was compiled by LER.

# Peptides as versatile scaffolds for quantum computing

Lorena E. Rosaleny,*,† Alicia Forment-Aliaga,*,† Helena Prima-García,*,† Ramón Torres-Cavanillas,† José J. Baldoví,†,§ Violetta Gołębiewska,† Karolina Wlazło,† Garin Escorcia-Ariza,† Luis Escalera-Moreno,† Sergio Tatay,† Carolina García-Llácer,† Miguel Clemente-León,† Salvador Cardona-Serra,† Patricia Casino,‡ Luis Martínez-Gil,‡ Alejandro Gaita-Ariño,*,† and Eugenio Coronado†

† *Instituto de Ciencia Molecular, Universitat de València, Catedrático José Beltrán 2, E46980 Paterna, Spain*

‡ *Dept. de Bioquímica y Biología Molecular, ERI BioTecMed, Universitat de València, Dr. Moliner 50, 46100 Burjassot, Spain*

§ *Nano-Bio Spectroscopy Group and European Theoretical Spectroscopy Facility (ETSF),*

*U. del País Vasco CFM CSIC-UPV/EHU-MPC & DIPC, Av. Tolosa 72, 20018 San Sebastián, Spain*

**E-mail:** lorena.rosaleny@uv.es; alicia.forment@uv.es; helena.prima@uv.es; alejandro.gaita@uv.es

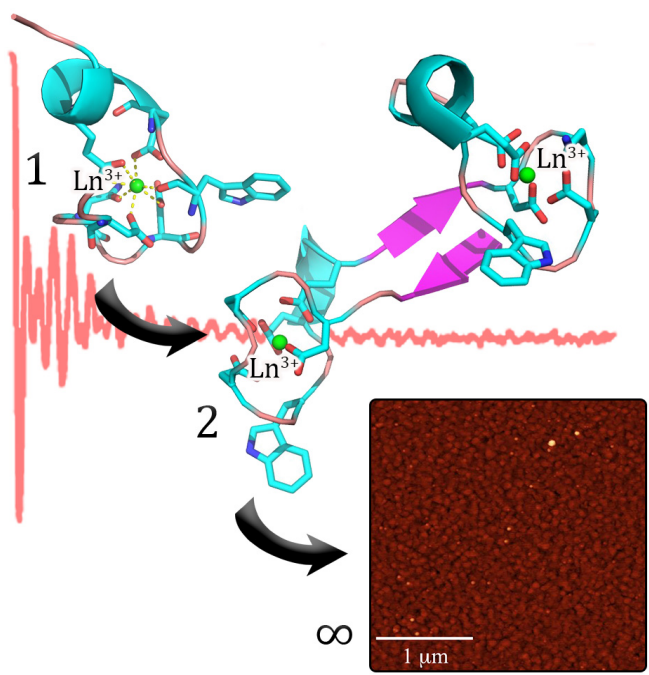

# List of contents

# S1 Spectroscopic characterization of LBT and LBTC peptides

We verified the purity of the **LBT** peptides by comparing the photophysical properties of **LBT** and **LBTC** with those reported in the literature.[1] The coordination ability of **LBT** and **LBTC** to lanthanides was proven by luminescence measurements of 5 μM 1:1 LBT:$Tb^{3+}$ and LBTC:$Tb^{3+}$ buffered solutions (see Figure S1). The emission spectra of tryptophan-sensitized $Tb^{3+}$ were measured using a PTI MO-5020 spectrofluorimeter in 1-cm path length quartz cells. Luminescence was measured at pH 7.0 in 10 mM N-2-hydroxyethylpiperazine-N'-2-ethanesulfonic acid (HEPES) buffer, 100 mM NaCl, and 5 mM 2-mercaptoethanol (**LBT** solutions) or 0.5 mM tris(2-carboxyethyl)phosphine (TCEP) (**LBTC** solutions). These solutions contained 5 μM **LBT** or **LBTC**, in the presence or absence of $Tb^{3+}$. The spectra were recorded from 300 to 600 nm by exciting at 280 nm. A 295-nm longpass filter to remove interference from harmonic doubling was used. The spectra showed the characteristic peaks of Lanthanide Binding Tags complexed with $Tb^{3+}$ at the same wavelengths as those reported in the literature (main peak at 544 nm).[1] Note that two characteristic **LBT** emmision bands are detected both in presence and absence of $Tb^{3+}$, a wide one around 360 nm and another, less wide, at 565 nm; the wide band is also present in the **LBTC** luminiscence. Binding to $Tb^{3+}$ produces three additional characteristic peaks, the main peak at 544 nm and two weaker ones, at 480 nm and 590 nm.

Additionallly, titration of **LBT** and **LBTC** solutions with $Tb^{3+}$ were performed (data not shown). The maximum emission is obtained with a 1:1 peptide:$Tb^{3+}$ ratio. Competitive titration with $Gd^{3+}$ demonstrates the coordination of LBT to other lanthanides since the characteristic emission of $Tb^{3+}$ disappears completely after the addition of a 6x excess of $Gd^{3+}$ (data not shown).

To complete the spectroscopic characterization, the peptide concentration of the **LBT** and **LBTC** solutions was confirmed by UV absorption at 280 nm in a 70 μM **LBT** or **LBTC** solution with 6 M guanidinium chloride and 0.02 M phosphate pH 7.0 buffer using known extinction coefficients of tryptophan ($\varepsilon_W$ = 5685 $M^{-1}cm^{-1}$) and tyrosine residues ($\varepsilon_Y$ = 1285 $M^{-1}cm^{-1}$).[2] This calculation assumes extended conformation for the peptide, as the measure is done in denaturing conditions because of the use of a chaotropic agent (guanidinium chloride). Absorption spectra were recorded on a Jasco V-670 spectrophotometer.

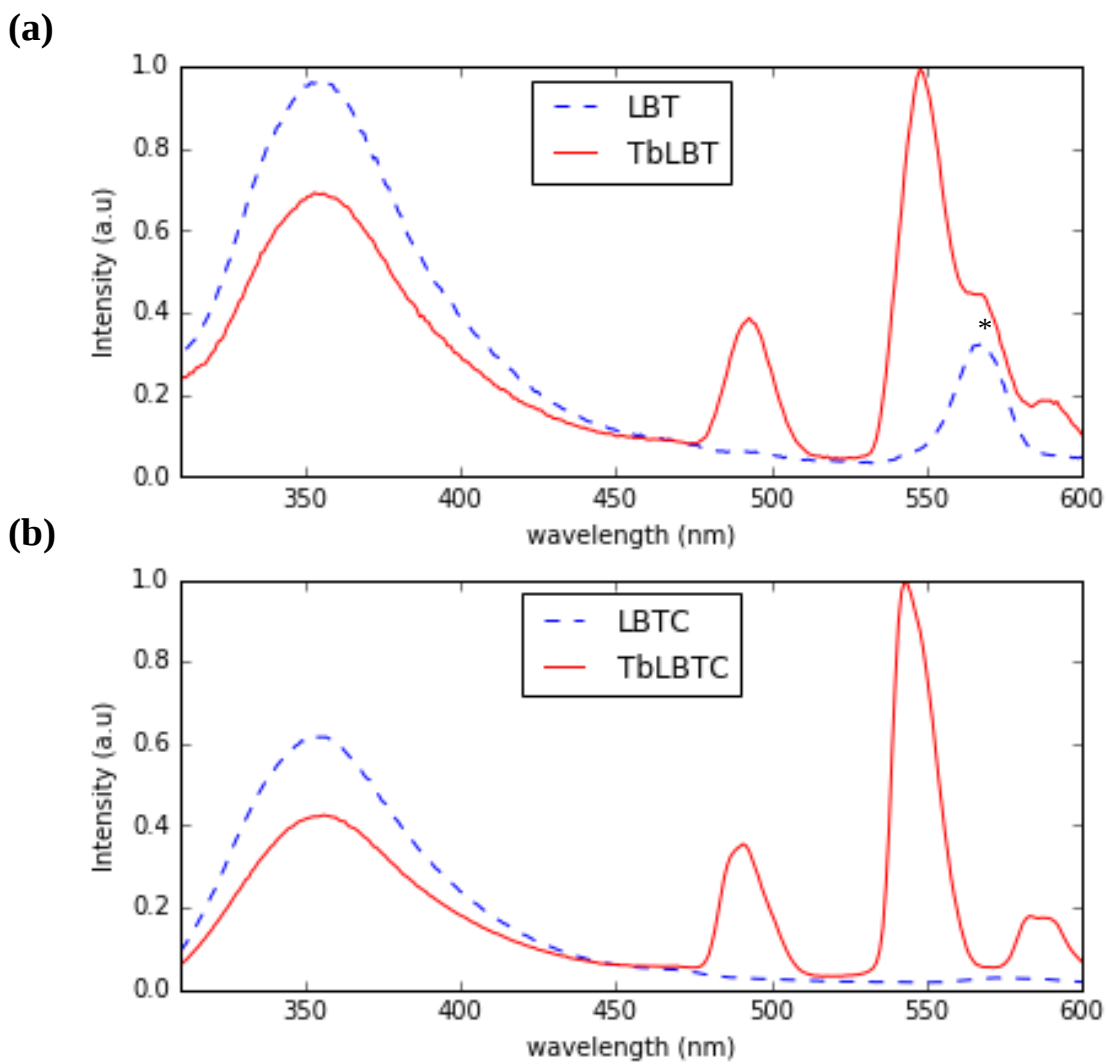

*Figure S1.* Luminiscence spectra (emission spectra upon 280 nm excitation) from (a) 5 μM **LBT** with/without 5 μM $Tb^{3+}$ (b) 5 μM **LBTC** with/without 5 μM $Tb^{3+}$. In all cases, solutions contain 10 mM HEPES pH 7.0 buffer and 100 mM NaCl. Note the additional 3 peaks presented by the Tb derivatives at about 490, 545 and 580, both as TbLBT and TbLBTC. (The LBT peak at 570 nm, and the corresponding shoulder for **TbLBT**, marked with a \*, are an experimental artifact due to an unfiltered optical harmonic).

# S2 Continous Wave (CW) EPR spectroscopy and pulsed EPR determination of relaxation times

EPR measurements were performed in buffer containing 10 mM HEPES at pH 7.0, 100 mM NaCl, 5 mM β-mercaptoethanol and 20 % glycerol (for cryoprotection). Thorough degassing was performed on all solutions previously to the EPR measurements, with a freeze-pump-thaw procedure as described in a work from Hirsh and Brudvig.[3] This is necessary since the presence of paramagnetic $O_2$ traces significantly alters pulsed EPR results in this kind of frozen samples.

The solutions measured by EPR spectroscopy contained varying concentrations of LBT peptide (from 72 μM to 144 μM) which were in excess respect to the complexing lanthanide ion.

Both continuous-wave and pulsed EPR data were recorded on an ELEXSYS E580 EPR spectrometer (Bruker) equipped with a Pulsed X-band (9.3 GHz cavity and resonators) operating in the range 4 -300 K with X band sources. The external magnetic field $B_0$ can be applied in a range 0-2 T. The pulsed X-Band is equipped with a TWT (travelling wave tube) amplifier, and is able to perform pulses between 0.7 ns and 15 µs long.

The pulse sequences for the echo-induced, $T_2$ and $T_1$ experiments was $T_{\pi/2}$=12 ns, $T_{\pi}$=24 ns with a fixed interpulse delay time of $\nu$ = 200 ns. Rabi oscillations were acquired by long microwaves field pulses of length t and subsequently recorded by spin echo. The t pulse and the π/2- π Hahn spin-echo sequence were recorded varying the attenuation power and subsequently $B_1$. The $B_0$ was 3480 G.

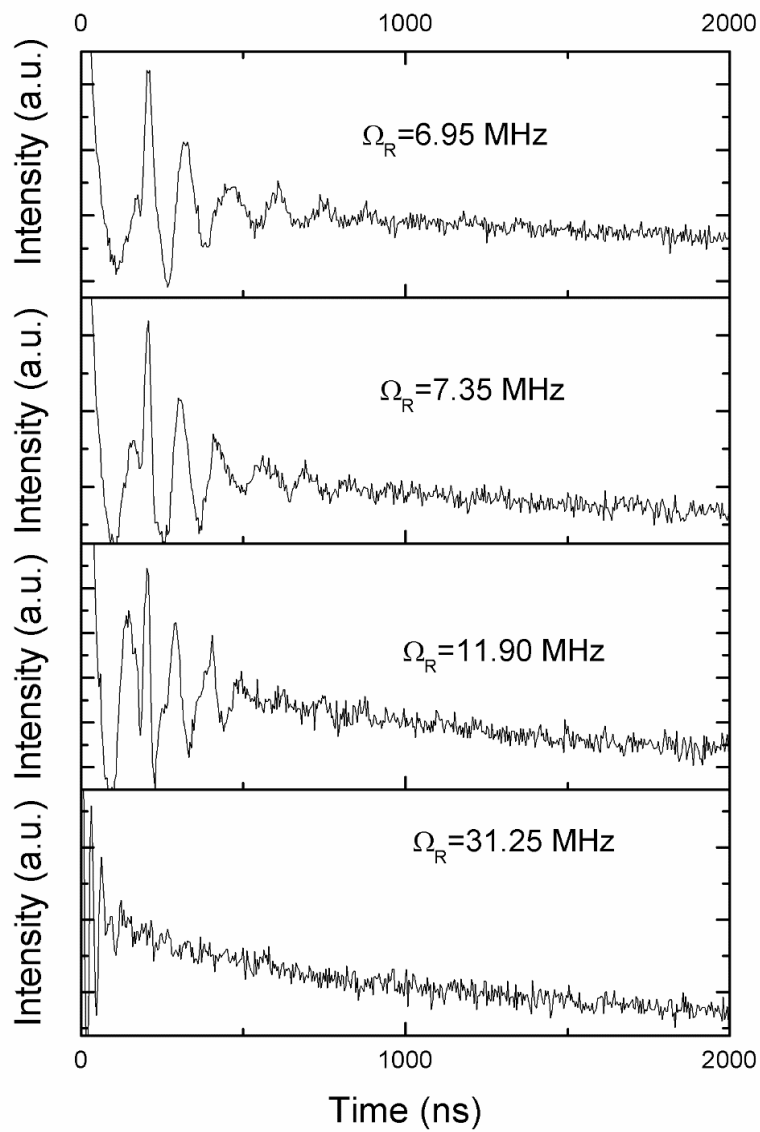

***Figure S2.1.*** *Measured $M_z(t)$ showing the result of nutation experiments at microwave powers that are above or below values where the Rabi frequency matches the Larmor frequency of the proton for the sample **GdLBT** at 4 K and at $H_0$= 3480 G*

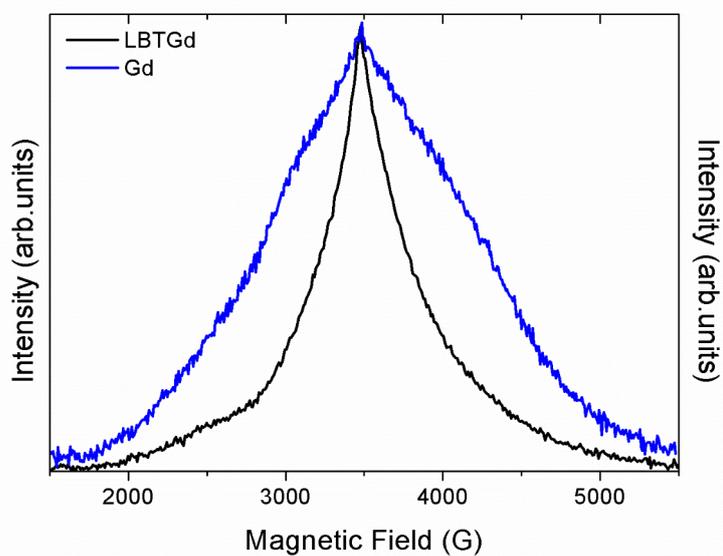

*Figure S2.2.* Band echo-detected EPR as a function of magnetic field for **GdLBT** (black line) and Gd without LBT (blue line) measured at 4.5 K. Pulses of 12 ns for π/2 and 24 ns for π.

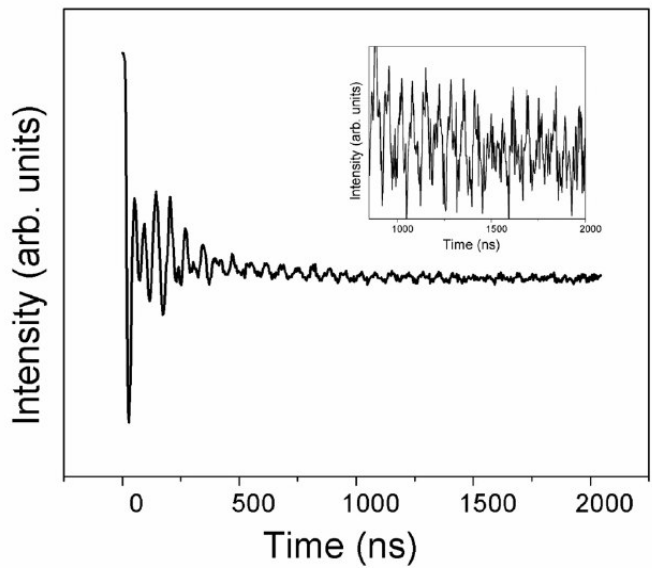

*Figure S2.3.* Rabi oscillations adjusting the microwave attenuation to achieve the Hartmann-Hahn condition, (i.e. the electronic Rabi frequency matches the Larmor frequency of the proton) for a Gd sample prepared without the LBT peptide, measured at 4 K and at $H_0$= 3480 G. Inset: zoom-in at Rabi oscillations for long times.

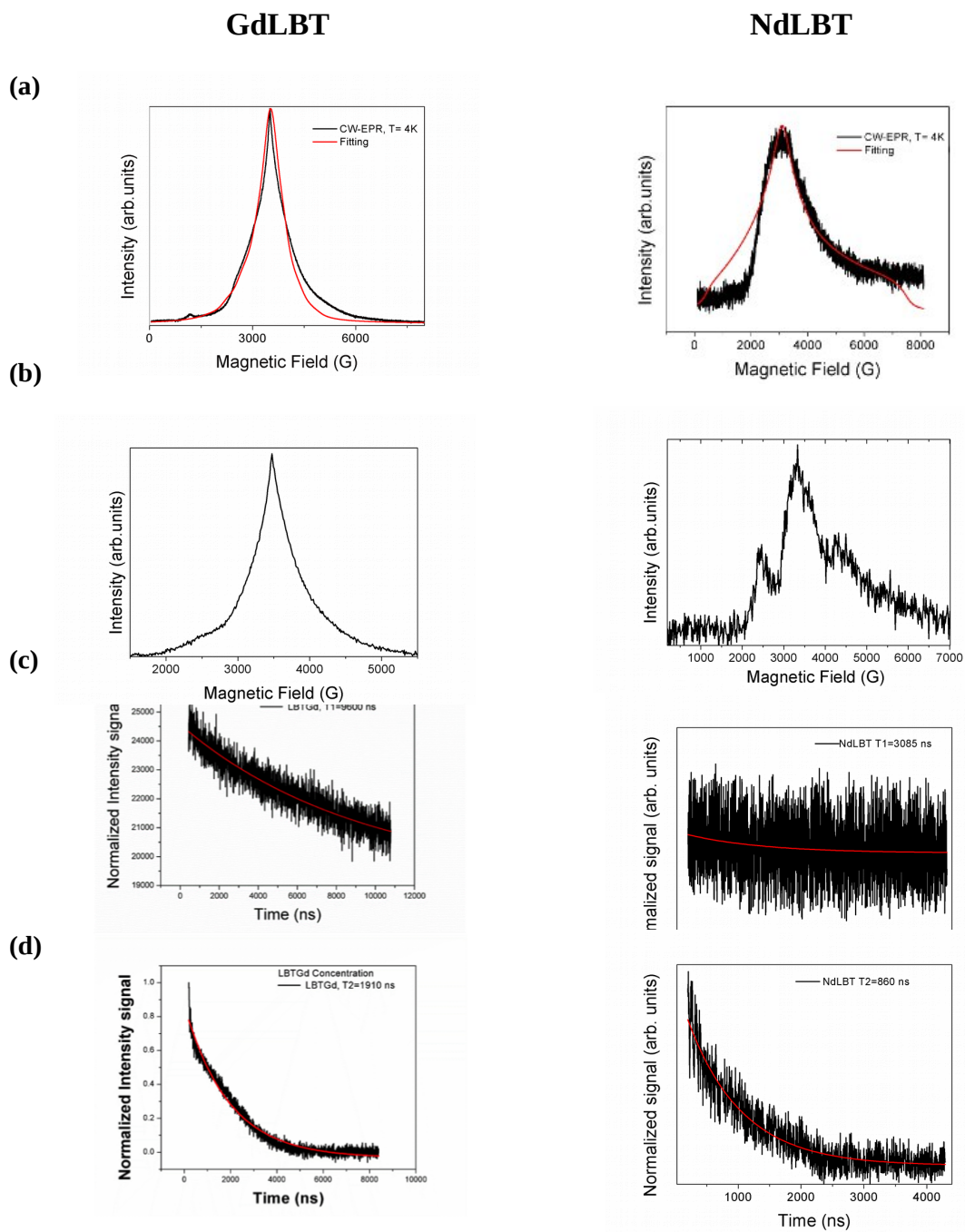

*Figure S2.4.* X-band (9.71 GHz) EPR experiments at 4K on a frozen solution of **LnLBT** (100 μM) which contained 20% Glycerol. Left: **GdLBT**, Right: **NdLBT**. (a) continuous wave EPR, (b) echo-detected EPR, (c) $T_1$, and (d) $T_2$ determination. Fitting curves are monoexponential functions with no stretch.

## S2.1 Spin energies and spin states

The energy-level scheme and the spin states of **GdLBT** (J = S = 7/2) and **NdLBT** (J = 9/2) in presence of a external static magnetic field B were extracted from the following spin hamiltonian:

$$\hat{H} = D\left[S_z^2 - \frac{1}{3}S(S+1)\right] + E\left(S_x^2 - S_y^2\right) + \mu_B g \vec{B} \cdot \vec{S}$$

The magnetic anisotropy parameters *D*, *E* and the Landé *g* factor, which was assumed to be isotropic, were extracted from fitting the cw-EPR spectrum (see figure S2.4.a left for **GdLBT** and right for **NdLBT**) by using the "pepper" package of the software EasySpin. The best fit parameters found for **GdLBT** are *D* = -640 MHz, *E* = 227 MHz, *g* = 1.9723, while for **NdLBT** are *D* = -233.333 GHz, *E* = 166.667 GHz, *g* = 2.00232. No strains for *D*, *E*, *g* were needed.

***Figure S2.5.*** *"Spin energies vs magnetic field" Zeeman diagram of **GdLBT** (S = J = 7/2) when the magnetic field is applied along the z axis. Resonant magnetic fields at X-band (9.71 GHz) for each transition with $\Delta m_J = \pm 1$, $\Delta m_I = 0$ are shown, being $m_J$ and $m_I$ the quantum numbers of the operators $J_z$ (electron spin) and $I_z$ (nuclear spin, I = 3/2), respectively. Each one of the eight curves is really 4-fold degenerate as no hyperfine coupling was set in the cw-EPR fitting.*

## S2.2 Nuclear spin bath decoherence: theoretical calculations

Decoherence calculations were performed by using the SIMPRE1.2[4] package, which was adapted to deal with randomly oriented samples (powder or solution). The energy interaction square between the qubit (electron spin) and the i$^{th}$ nucleus of the bath, due to the interaction energy difference when the qubit is in its symmetric, *S*, and antisymmetric, *A*, state, is given by:

$$\left\|E_i^A - E_i^S\right\|^2 =$$

$$\left(\frac{\mu_0 \mu_B \mu_N g_N^i}{8\pi}\right)^2 \frac{1}{r_i^6} \sum_{\alpha=x,y,z}\left(\langle \hat{I}_\alpha\rangle_i\left(\left(1-3\left(\frac{(\vec{r}_i)_\alpha}{r_i}\right)^2\right)g_\alpha\left(\langle \hat{J}_\alpha\rangle^A - \langle \hat{J}_\alpha\rangle^S\right) - 3\frac{(\vec{r}_i)_\alpha}{r_i^2}\sum_{\substack{\beta=x,y,z \\ \beta\neq\alpha}}(\vec{r}_i)_\beta g_\beta\left(\langle \hat{J}_\beta\rangle^A - \langle \hat{J}_\beta\rangle^S\right)\right)\right)^2$$

being $\mu_0$ the magnetic constant, $\mu_B$ the Bohr magneton, $\mu_N$ the nuclear magneton, $g_N^i$ the Landé *g* factor of the i$^{th}$ nucleus, $r_i$ the distance between the magnetic ion (qubit) and the i$^{th}$ nucleus and $g_\alpha$ is the $\alpha$ = x,y,z component of the electron *g* tensor (they will coincide with the *g* tensor main values if the magnetic axes coincide with the canonical cartesian basis set, such as we assume in the EasySpin fit and in our calculations). The expectation values of the x,y,z components of the electron spin *J* operator when the qubit is in its symmetric and antisymmetric state are provided by SIMPRE1.2. The same expectation values for the nuclear spin *I* operator are given by $(I_i(I_i+1)/3)^{1/2}$ in the high-temperature regime which can be assumed from typical low temperatures of 3-4 K in common EPR experiments, being $I_i$ the nuclear spin of the i$^{th}$ nucleus.

To adapt the above expression for randomly oriented samples (powder or solution) one has to expand this expression and average the director cosinus over the unitary sphere:

$$\gamma_i^\alpha := \cos\theta_i^\alpha = \frac{(\vec{r_i})_\alpha}{r_i} \quad ; \quad \left\langle (\gamma_i^\alpha)^n \right\rangle = \frac{1}{4\pi} \int_0^{2\pi} \left( \int_0^\pi \cos^n\theta_i^\alpha \sin\theta_i^\alpha d\theta_i^\alpha \right) d\phi_i^\alpha \quad ; \quad n \in \mathbb{N}$$

The values we need are $n = 1, 2, 3, 4$; giving 0, 1/3, 0, 1/5, respectively, for each $\alpha$ = x,y,z and for each nucleus i. The average energy interaction square between the qubit (electron spin) and the $i^{th}$ nucleus of the bath is now:

$$\overline{\left\| E_i^A - E_i^S \right\|^2} =$$

$$\left( \frac{\mu_0 \mu_B \mu_N g_N^i}{8\pi} \right)^2 \frac{1}{r_i^6} \sum_{\alpha=x,y,z} \left\langle \hat{I}_\alpha \right\rangle_i^2 \left( \frac{4}{5} g_\alpha^2 \left( \left\langle \hat{J}_\alpha \right\rangle^A - \left\langle \hat{J}_\alpha \right\rangle^S \right)^2 + \sum_{\substack{\beta=x,y,z \\ \beta \neq \alpha}} g_\beta^2 \left( \left\langle \hat{J}_\beta \right\rangle^A - \left\langle \hat{J}_\beta \right\rangle^S \right)^2 \right)$$

The nuclear spin bath contribution to the total half-width square is given by:

In the high-field regime, meaning $\Delta \gg E_n^2$, being $\Delta$ the qubit gap, which $$E_n^2 = \sum_i \overline{\left\| E_i^A - E_i^S \right\|^2}$$

is a condition widely fulfilled in common EPR experiments at X-band, the qubit quantum dynamics in presence of a nuclear spin bath can be solved perturbatively and both the nuclear decoherence time and its associated decoherence rate are:

$$T_2^n = \hbar \frac{\Delta}{E_n^2} \qquad \gamma_n = 2 \frac{E_n^2}{\Delta^2}$$

To simulate the solution, the **GdLBT** molecule was placed inside a sphere of a given radius $R$ with the $Gd^{3+}$ ion right at the center. We take only water as solvent, considering that glycerol molecules have a very similar proton content compared to water molecules. We need to do this approximation because the model we employ to simulate the solution considers the solvent molecules as spheres with a certain radius $r$ (a water molecule is closer to be a sphere than a glycerol molecule). Now, we fill another empty sphere of radius $R$ only with solvent water molecules. This is done as follows: fixing a maximum number $M$ of attempts, the code places the first sphere of radius $r$ on a random position inside the sphere of radius $R$. Then, a second sphere of radius $r$ is placed in another random position inside the sphere of radius $R$. If the centers of these two spheres are closer than $2*r$, the second sphere is rejected and the code finds another random position for the second sphere until it does not overlap with the first one, meaning their centers are further away than $2*r$. When the number of attemps trying to find a random position for the second sphere reaches $M$, the code is stopped. If finding a random place for the second sphere has succeeded, then the code moves to find a random place for a third sphere of radius $r$, checking now the overlapping with the first and the second sphere.

The process is repeated until a sphere of radius $r$ cannot be placed within $M$ attempts. To assure that there are not hollows too large left between solvent molecules, the last successful insertion of a sphere of radius $r$ before stopping the code must have been done at an attempt $m$ much lower than $M$. If $m$ is close to $M$, the number $M$ should then be increased and the code run again.

Several code executions are performed and the radius $r$ is adjusted until the water density calculated inside the sphere matches the water density at a given temperature (4K in our case). Once this is done, we keep this effective $r$ radius and go back to the very first sphere of radius $R$ where we placed the **GdLBT** molecule. Here, we consider that each atom of this molecule is a sphere of radius equal to its Van der Waals radius. Now, we start the same procedure described above to fill the sphere of radius $R$ with water molecules being each one of them spheres of radius $r$. The difference now is that the condition for the first sphere of radius $r$ to be placed successfully is that it must not overlap with any atom of the **GdLBT** molecule, and so on for the next ones. Finally, assuming that the solvent water molecules are far away enough from the magnetic ion (qubit) so that distinguishing the geometrical structure of the water molecule is not important, we place two hydrogen atoms and one oxygen atom right at the center of each sphere of radius $r$. This is the nuclear spin bath that we take as an input for SIMPRE1.2. The radius $R$ was increased from 40 to 65 Å until $T_2^n$ was converged. We selected $R$ = 60 Å. The maximum number of attemps was 10.000.000 and the effective radius $r$ was 1.395 Å, which simulates a water density of 0.925 g/cm$^3$ at 16 K or below this temperature. The $T_2^n$ calculation was performed at a static magnetic field of 0.34 T applied along the z direction, giving a qubit gap of $\Delta$ = 9.60 GHz, typical of X-band EPR experiments.

The resonant transition selected, i.e., the two spin states of the qubit selected, was:

$|m_J = -1/2, m_I = +3/2\rangle \rightarrow |m_J = +1/2, m_I = +3/2\rangle$.

There are still three more degenerate resonant transitions since no hyperfine coupling was set in the EasySpin fit. These degenerate resonant transitions in the decoherence calculations were:

$|m_J = -1/2, m_I = +1/2\rangle \rightarrow |m_J = +1/2, m_I = +1/2\rangle$,

$|m_J = -1/2, m_I = -1/2\rangle \rightarrow |m_J = +1/2, m_I = -1/2\rangle$,

$|m_J = -1/2, m_I = -3/2\rangle \rightarrow |m_J = +1/2, m_I = -3/2\rangle$,

which gave similar nuclear decoherence times $T_2^n$.

# S3 Steps for the design of a low spin-vibrational-coupling peptidic spin

A major advantage of using peptidic spin qubits as building blocks for spintronics is the possibility for systematic property optimization: whether one is interested in quantum coherence or in any other specific spintronic effect, the rational procedure to improve the desired property would be to perform a screening of a combinatorial peptide library, i.e. to make combinations of substitutions of individual aminoacids in the sequence to obtain progressively better properties.

In this case, we suggest one would proceed as follows:

In a first phase, on would employ an inexpensive approach such as Force Fields (e.g. using the Amber suite) to estimate the molecular vibrations of a given peptidic spin qubit. With a set of normal vibrational modes, it is then possible to generate a set of distorted geometries, following each of these normal modes. One would perform an energy cutoff discarding all vibrational modes above 300 K and thus greatly simplifying the problem, since there is no interest in calculating the high-energy modes which will not be populated at any reasonable temperatures. One would then calculate the energy of the spin qubit, corresponding in our case to the magnetic levels of the ground NdLBT doublet at the operating magnetic field. This can be done inexpensively employing an effective electrostatic approach such as the REC model with SIMPRE or with more care by ab initio methods (MOLCAS or ORCA). As presented recently in Escalera-Moreno *et al.*[5] this is enough to determine the set of key local vibrations in the relaxation of a molecular spin qubit (or a single-molecule magnet).

The second phase of the work would involve designing a peptidic library, based on the original peptidic spin qubit. From the first phase one can determine, for the vibrations that are key for spin relaxations, which are the aminoacids with the greatest displacement. Moreover, one can also determine which are the aminoacid-aminoacid distances that are most affected by these key vibrations. Exploring controlled changes in precisely these two sets of aminoacids should
guide the design of the peptidic library, which can then be systematically explored. From the point of view of the experiment and to save time, we suggest that it is enough for each peptide to do a relatively inexpensive test by means of an quick EPR experiment, such as determining the presence and intensity of a spin echo at 100K, which does not require the time- (and resource-) intensive task of cooling the system to liquid He temperatures. This would allow exploring many samples a day.

## S4 Magnetometry measurements (dc)

Magnetic susceptibility measurements were performed on a Quantum Design MPMS-XL-5 magnetometer equipped with a SQUID sensor. The samples were prepared by lyophilization overnight of an aqueous solution of **LnLBT** containing 1 mM LBT, 0.66 mM $LnCl_3$, 100 mM NaCl and 10 mM HEPES pH 7.0 with a bencthop manifold freeze-drier. To guarantee a quantitative coordination of the Ln, an excess of LBT was employed.

The susceptibility data were corrected from the diamagnetic contributions as deduced by using Pascal's constant tables. Due to experimental difficulties we only acquired direct current (dc) data in the range 2-30 K with an applied field of 1000 G, and only for **TbLBT** and **DyLBT** (and **NdLBT**).

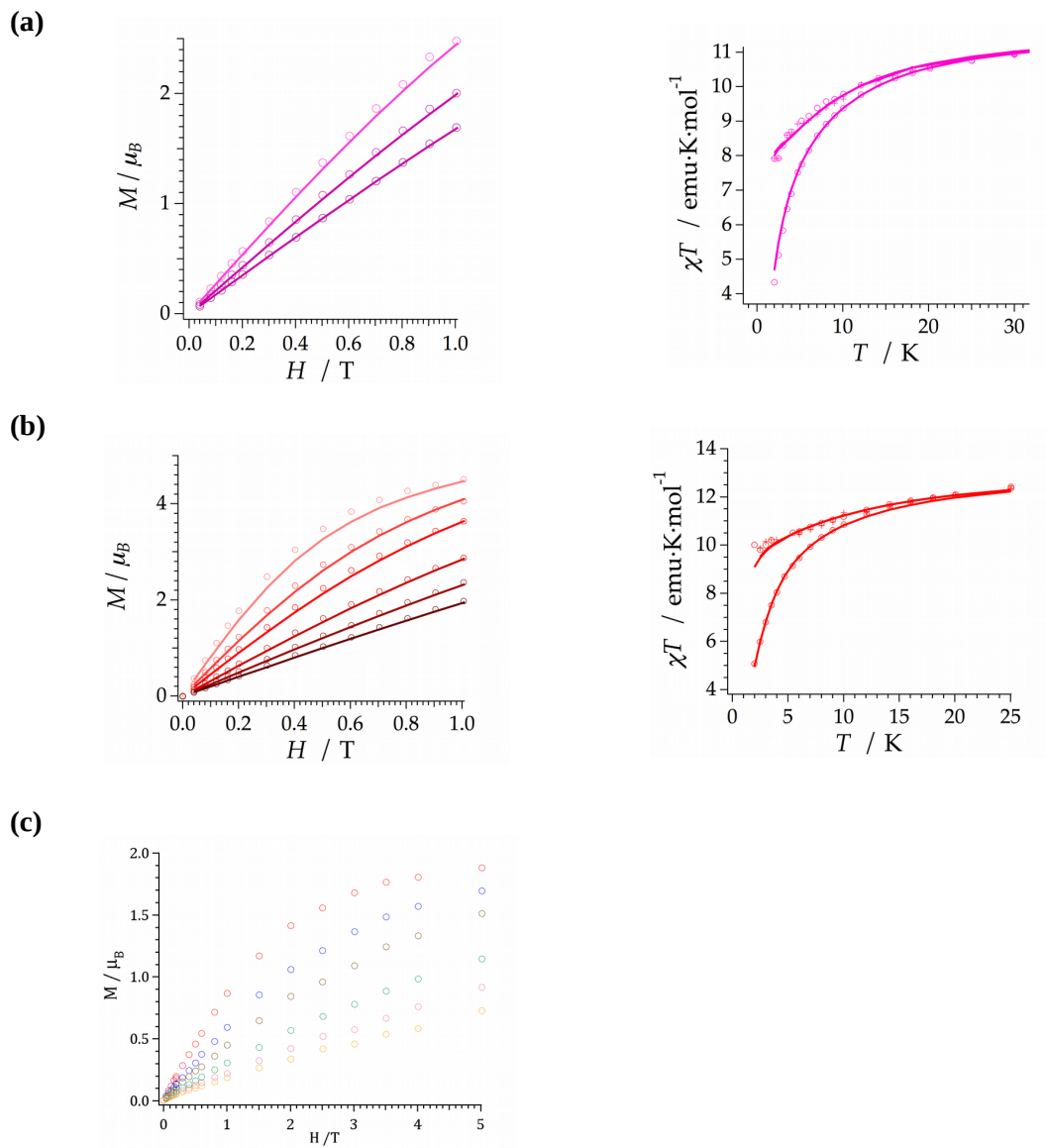

*Figure S4.* Left: magnetization M vs H, Right: magnetic susceptibility $\chi T$ vs T. (a) **TbLBT**, (b) **DyLBT**, (c) **NdLBT**. Circles: experimental data. Continuous line: theoretical fit, see section S5.

# S5 Theoretical fit of dc magnetometry employing the CONDON package

Magnetic susceptibility and magnetization data of **TbLBT** and **DyLBT**, as presented in section S4 were fitted employing a phenomenological, effective Hamiltonian, containing only diagonal terms $B_{20}$, $B_{40}$, $B_{60}$ and an extradiagonal component Re$B_{44}$, Im$B_{44}$. The complexity of the geometry prevented our use of SIMPRE1.2,[4] so we used the CONDON package instead.[6]

**Table S5.1.** *Parameters of the Ligand-Field Hamiltonian (cm$^{-1}$) in the Wybourne notation for the **LnLBT** complexes. For Ln = Tb, Dy, the LFPs are fitted and error bars are shown between parenthesis.*

|            | **Tb**      | **Dy**      |
|------------|-------------|-------------|
| $B_{20}$   | 368 (25)    | -70 (6)     |
| $B_{40}$   | 654 (11)    | 982 (20)    |
| Re$B_{44}$ | -180 (8)    | -765 (41)   |
| Im$B_{44}$ | -468 (18)   | 3 (57)      |
| $B_{60}$   | 595 (121)   | -716 (33)   |

**Table S5.2.** *Amplitude of $|M_J\rangle$ contributing to the ground state of **LnLBT**.*

| Tb | 94% |-5⟩ |
|----|--------------------------|
| Dy | 52% |-3/2⟩ + 45% |+5/2⟩ |

# S6 Determination of folding patterns by means of the program SHAPE

SHAPE software quantifies the deviation of chemical 3D structures from either ideal geometrical structures or selected structures from the user.[7] In our case, we used it to compare the shape of LBT-containing peptides previously described in the literature with each other (see table 1 in main text). SHAPE software determines a continuous shape parameter (SP), which is mathematically defined in a way that is independent of the size of the system. By definition, the resulting value of SP is zero when the real coordinates of the metal site (problem structure, P) show exactly the desired ideal shape, and increases with the degree of distortion of the structure. Values below 0.1 represent chemically insignificant distortions in the structure. Values larger than 3 mean important distortions, with the highest values commonly encountered being in the order of 40.

The main object for comparison were the α-carbons forming the backbone of each of the LBT-containing peptides, namely LBT peptide (molecule A and B, PDB id 1TJB), dLBT-ubiquitin (molecule A and B, PDB id 2OJR), Interleukin-1β-(S1-LBT) (PDB id 3LTQ), xq-dSE3-ubiquitin (molecule A.1, A.2, B.1 and B.2, PDB id 3VDZ) and Z-L2LBT (PDB id 2LR2). The coordinates for all of the proteins were extracted from the PDB files downloaded from the RCSB Protein Data Bank website (https://www.rcsb.org/pdb/).

It is observed that the aminoacidic sequence is highly similar from the first fifteen α-carbon atoms belonging to the LBT motif in all of the examined proteins, and it is much more variable in the sixteenth and seventeenth α-carbon atoms (table S6.1). Additionally, not all of the LBT motifs in the molecules contain seventeen α-carbons. For this reason two different sets of calculations were carried out. Regarding the first set, only fifteen α-carbons were taken into account (table S6.2), whereas the second one included seventeen α-carbons when possible (table S6.3).

*Table S6.1.* *Aminoacid sequence of the LBT motifs found in different chimeric polypeptides. For double LBT motifs the notation used the chain name in the PDB files (A or B) and a number to denote if the unit was first or second in the sequence.*

| LBT motif (PDB ID) | Cα position in motif | | | | | | | | | | | | | | | | |
|---|---|---|---|---|---|---|---|---|---|---|---|---|---|---|---|---|---|
| | 1CA | 2CA | 3CA | 4CA | 5CA | 6CA | 7CA | 8CA | 9CA | 10CA | 11CA | 12CA | 13CA | 14CA | 15CA | 16CA | 17CA |
| 1TJB chainA&B | TYR | ILE | ASP | THR | ASN | ASN | ASP | GLY | TRP | TYR | GLU | GLY | ASP | GLU | LEU | LEU | ALA |
| 2OJR chainA.1 | TYR | ILE | ASP | THR | ASN | ASN | ASP | GLY | TRP | ILE | GLU | GLY | ASP | GLU | LEU | - | - |
| 2OJR chainA.2 | TYR | ILE | ASP | THR | ASN | ASN | ASP | GLY | TRP | ILE | GLU | GLY | ASP | GLU | LEU | LEU | ALA |
| 3LTQ | TYR | ILE | ASP | THR | ASN | ASN | ASP | GLY | TRP | ILE | GLU | GLY | ASP | GLU | LEU | TYR | ASP |
| 3VDZ chainA.1&B.1 | TYR | ILE | ASP | THR | ASP | ASN | ASP | GLY | SER | ASP | GLY | ASP | GLU | LEU | - | - | - |
| 3VDZ chainA.2&B.2 | TYR | ILE | ASP | THR | ASP | ASN | ASP | GLY | SER | ILE | ASP | GLY | ASP | GLU | LEU | LEU | ALA |
| 2LR2 | TYR | ILE | ASP | THR | ASN | ASN | ASP | GLY | ALA | TYR | GLU | GLY | ASP | GLU | LEU | SER | GLY |

**Table S6.2.** *SHAPE results calculated using 15 α-carbons from the LBT motifs.*

|               | 1TJB chainA | 1TJB chainB | 2OJR chainA.1 | 2OJR chainA.2 | 3LTQ  | 3VDZ chainA.1 | 3VDZ chainA.2 | 3VDZ chainB.1 | 3VDZ chainB.2 | 2LR2  |
|---------------|-------------|-------------|---------------|---------------|-------|---------------|---------------|---------------|---------------|-------|
| 1TJB chainA   | 0.000       | 0.147       | 3.174         | 2.312         | 3.113 | 4.092         | 2.135         | 4.243         | 2.116         | 0.260 |
| 1TJB chainB   | 0.147       | 0.000       | 2.553         | 1.829         | 2.409 | 3.665         | 1.787         | 3.715         | 1.783         | 0.206 |
| 2OJR chainA.1 | 3.174       | 2.553       | 0.000         | 1.589         | 0.302 | 0.737         | 2.162         | 0.685         | 2.169         | 3.161 |
| 2OJR chainA.2 | 2.312       | 1.829       | 1.589         | 0.000         | 1.481 | 2.258         | 0.551         | 2.114         | 0.591         | 1.998 |
| 3LTQ          | 3.113       | 2.409       | 0.302         | 1.481         | 0.000 | 1.040         | 1.927         | 0.909         | 1.919         | 3.171 |
| 3VDZ chainA.1 | 4.092       | 3.665       | 0.737         | 2.258         | 1.040 | 0.000         | 3.047         | 0.243         | 3.007         | 4.276 |
| 3VDZ chainA.2 | 2.135       | 1.787       | 2.162         | 0.551         | 1.927 | 3.047         | 0.000         | 2.849         | 0.013         | 1.897 |
| 3VDZ chainB.1 | 4.243       | 3.715       | 0.685         | 2.114         | 0.909 | 0.243         | 2.849         | 0.000         | 2.830         | 4.329 |
| 3VDZ chainB.2 | 2.116       | 1.783       | 2.169         | 0.591         | 1.919 | 3.007         | 0.013         | 2.830         | 0.000         | 1.903 |
| 2LR2          | 0.260       | 0.206       | 3.161         | 1.998         | 3.171 | 4.276         | 1.897         | 4.329         | 1.903         | 0.000 |

**Table S6.3.** *SHAPE results calculated using 17 α-carbons from the LBT motifs.*

|               | 1TJB chainA | 1TJB chainB | 2OJR chainA.2 | 3LTQ  | 3VDZ chainA.2 | 3VDZ chainB.2 | 2LR2  |
|---------------|-------------|-------------|---------------|-------|---------------|---------------|-------|
| 1TJB chainA   | 0.000       | 0.409       | 9.343         | 6.180 | 8.095         | 8.321         | 2.839 |
| 1TJB chainB   | 0.409       | 0.000       | 7.530         | 3.956 | 6.801         | 7.023         | 1.477 |
| 2OJR chainA.2 | 9.343       | 7.530       | 0.000         | 7.595 | 0.648         | 0.681         | 5.749 |
| 3LTQ          | 6.180       | 3.956       | 7.595         | 0.000 | 8.342         | 8.540         | 3.570 |
| 3VDZ chainA.2 | 8.095       | 6.801       | 0.648         | 8.342 | 0.000         | 0.014         | 5.753 |
| 3VDZ chainB.2 | 8.321       | 7.023       | 0.681         | 8.540 | 0.014         | 0.000         | 5.964 |
| 2LR2          | 2.839       | 1.477       | 5.749         | 3.570 | 5.753         | 5.964         | 0.000 |

**Table S6.4.** *SHAPE results calculated using oxygen atoms from the LBT motifs coordination environment.*

|             | 1TJB chainA | 1TJB chainB | 2OJR 15CA | 2OJR 17CA | 3LTQ  |
|-------------|-------------|-------------|-----------|-----------|-------|
| 1TJB chainA | 0.000       | 0.668       | 1.962     | 4.229     | 0.920 |
| 1TJB chainB | 0.668       | 0.000       | 1.608     | 3.622     | 0.367 |
| 2OJR 15CA   | 1.962       | 1.608       | 0.000     | 1.521     | 1.633 |
| 2OJR 17CA   | 4.229       | 3.622       | 1.521     | 0.000     | 3.442 |
| 3LTQ        | 0.920       | 0.367       | 1.633     | 3.442     | 0.000 |

The oxygen atoms forming the coordination environment for each of the LBT motifs were used for SHAPE calculations presented in table S6.4.

In sum, the analyses of the information of the tables above is the following:

-for the 15-long peptide fragment (table S6.2), the foldings are distorted compared with each other, but remarkably similar, considering the great difference between the aminoacid sequences, for example in the case of 1TJB vs 3VDZ:

-for the 17-long peptide fragment (table S6.3), the folding is markedly more different, signaling that the last two aminoacids do not fold reproducibly: the maximum SP parameters double compared with the previous table: from slightly above 4 to slightly above 8

-for the 8 oxygen coordination environment (table S6.4), the shapes are distorted but again they are remarkably similar, considering the relative freedom enjoyed by the side groups (again, the maximum SP parameters are slightly above 4); here note that 3VDZ was designed for an efficient interchange of water molecules directly coordinated to the Gd, and as a consequence its coordination environment is severely distorted (although, as seen in the tables above, the folding remains essentially the same), so we did not consider it for the comparison.

Analysing Table S6.4, it is evident that there are relatively large calculated distortions between the oxygen coordination spheres of 2OJR (both 15CA and 17CA) and 3LTQ, something striking since they share the aminoacid sequence (primary structure) of the region that actually binds to the lanthanide ion, i.e. YIDTNNDGWIEGDEL. Compared with regular distortions between crystallographically inequivalent copies of identical coordination complexes, these SP parameters are rather high, which could mean that the coordination environments are difficult to reproduce. However, here it is crucial to put these SHAPE parameters in perspective, and for that one needs to note that peptide crystallography involves an uncertainty in the atomic positions that is higher than usual in coordination chemistry. For this purposes we can take as a measure the Diffraction Precision Index (DPI) as presented by Gurusaran and coworkers.[7] We employed the data analysis software accessible at the server http://cluster.physics.iisc.ernet.in/dpi/ and, as seen in Table S6.5, found that the DPI values of the peptides under study in this section range from about 0.1 to almost 0.4 Å, which means the SP parameters we calculate are in this case strongly affected by crystallographic noise. Thus, these numbers should be understood as an upper limit to the real structural difference between the peptidic foldings, in the case of Tables S6.2 and S6.3, or between the coordination environments, in the case of Table S6.4. Further studies would be needed to more accurately quantify the maximum possible distortion between the coordination spheres of two lanthanide binding tags with the same primary sequence.

*Table S6.5. Resolution and DPI regarding the protein structures studied.*

| PDB ID | Resolution (Å) | Diffraction Precision Index DPI (Å) |
| --- | --- | --- |
| 1TJB | 2.0 | 0.196 |
| 2OJR | 2.6 | 0.329 |
| 3LTQ | 2.1 | 0.097 |
| 3POK | 1.7 | 0.077 |
| 3VDZ | 2.4 | 0.392 |

Additionally, a study of the geometry of octa-coordinated $Tb^{3+}$ in the LBT peptide (1TJB) context was performed comparing the position of coordination oxygens to vertices in reference polyhedra predefined in the SHAPE package. The results (see table S6.6) suggested that the LBT coordination environment does not adopt an ideal geometry, but is closely related to a square antiprism (SHAPE parameter approx. 3.6), a triangular dodecahedron (approx. 3.3) or a biaugmented trigonal prism (approx. 3.4). Note however that these descriptions are limited in precision for the above stated reasons of crystallographic uncertainty.

*Table S6.6.* *Geometry study of octa-coordinated lanthanide ions in an LBT unit (1TJB). SP parameters are shown. Comparisons were made with all 13 ideal geometries included as reference polyhedra for octa-coordinated metals in the SHAPE software.*

| Symmetry | Geometry | 1TJB chainA | 1TJB chainB |
|---|---|---|---|
| $D_{8h}$ | Octagon | 30.016 | 30.640 |
| $C_{7v}$ | Heptagonal pyramid | 22.975 | 22.496 |
| $D_{6h}$ | Hexagonal bipyramid | 14.981 | 15.752 |
| $O_h$ | Cube | 10.736 | 11.304 |
| $D_{4d}$ | Square antiprism | 3.645 | 2.622 |
| $D_{2d}$ | Triangular dodecahedron | 3.627 | 3.021 |
| $D_{2d}$ | Johnson - Gyrobifastigium (J26) | 12.709 | 13.832 |
| $D_{3h}$ | Johnson - Elongated triangular bipyramid (J14) | 26.491 | 27.036 |
| $C_{2v}$ | Johnson - Biaugmented trigonal prism (J50) | 4.538 | 4.176 |
| $C_{2v}$ | Biaugmented trigonal prism | 3.842 | 3.081 |
| $D_{2d}$ | Snub diphenoid (J84) | 5.341 | 5.628 |
| $T_d$ | Triakis tetrahedron | 11.282 | 12.122 |
| $D_{3h}$ | Elongated trigonal bipyramid | 21.773 | 23.001 |

# S7 Distance and relative orientation of two lanthanide coordination centers in daLBT

For the prediction of the distance and relative orientation between two given lanthanide coordination centers in peptides coordinating two or more ions such as **daLBT**, it is useful to analyse the two existing crystal structures of double lanthanide binding tag structures: 2OJR and 3VDZ, in both cases part of a modified ubiquitin protein. In the case of 2OJR, the lanthanide-binding part is at the N-terminal end of ubiquitin, while in 3VDZ it is inserted in a loop. Additionally, the two sequences have an important difference which in principle can affect the folding, namely the Trp residue in 2OJR is replaced with Ser in the case of 3VDZ. Other than that, in both cases, these are peptides consisting on the repetition of two identical lanthanide-binding sub-units with no spacer, like the **daLBT** we present in this work. Let us study their differences in terms of distance and relative orientation, since this is an important question for the organization of spin qubits.

In terms of the metal-metal distance, analysis of the structures shows that it is 19.1 Å in the case of 2OJR and 21.1 Å in case of 3VDZ (see Table S7). This probably means that non-spaced LBTs, such as **daLBT**, can be consistently expected to situate lanthanides at minimum distances of about 20 ± 1 Å. Longer distances can be trivially achieved, either, as described in section S10, by introducing the different lanthanide binding tag in loops at known distant positions in structurally characterized proteins, or by introducing bulky spacers, e.g. a poly-alanine segment which will tend to form a long $\alpha$-helix.

**Table S7.** Distances (metal-metal), and angles between first metal, $\alpha$-carbon belonging to Tyr in the first LBT unit, and second metal, and between first metal, $\alpha$-carbon belonging to Tyr in the second LBT unit, and second metal. Measurements on 2OJR and 3VDZ protein structures were performed using PyMOL.[9]

|  | $r(Ln_1-Ln_2)$ (Å) | $\alpha(Ln_1-C_\alpha(Tyr1)-Ln_2)$ (°) | $\alpha(Ln_1-C_\alpha(Tyr2)-Ln_2)$ (°) |
|---|---|---|---|
| **2OJR** | 19.1 | 133.8 | 119.7 |
| **3VDZ** | 21.1 | 124.7 | 138.5 |

A more challenging problem is the control of the relative orientation of two given lanthanide binding tags. In the case of 2OJR and 3VDZ, we took as reference points the position of the two metals and of the two $\alpha$-carbons belonging to Tyr aminoacids. One of these Tyr is at the beginning of the first sub-unit and the other one is precisely between the two sub-units, so we measured the angle formed by $Ln_1-C_\alpha(Tyr)-Ln_2$ as a rough estimate of the relative orientation of the two sub-units. As can be seen in table S7, these $Ln_1-C_\alpha(Tyr)-Ln_2$ angles are in the same range, but with important variations. As can be seen in the Figure S7, local interactions (a small $\beta$-sheet can be appreciated) seem to determine the relative orientations of the two sub-units, rather than the bulkier protein they are attached to. This permits speculating that some sort of control of the orientation might be attained through careful proteic design.

**(a)** **(b)**

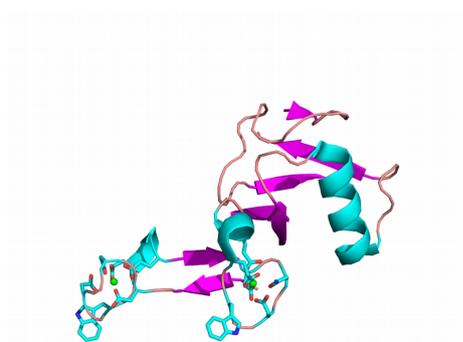 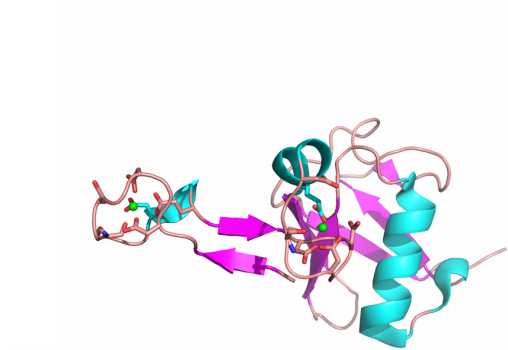

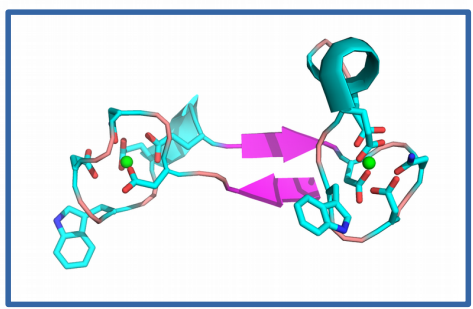 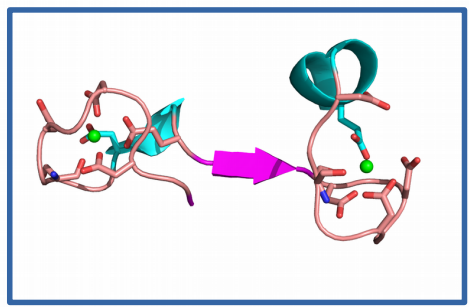

***Figure S7.** Structures from dLBT-ubiquitin (PDB id: 2OJR)[10] (a) and xq-dSE3-ubiquitin (PDB id: 3VDZ)[11] (b) depicted using PyMOL are shown. The magenta arrow represents β-sheet secondary structure, whereas the light blue helix shows a $3_{10}$ helix. The structures shown enclosed in blue rectangles display only the dLBT fragment belonging to dLBT-ubiquitin and xq-dSE3-ubiquitin respectively. The aminoacidic sequences corresponding to the fragments shown are: PGYIDTNNDGWIEGDELYIDTNNDGWIEGDELLA (belonging to dLBT-ubiquitin), and YIDTDNDGSIDGDELYIDTDNDGSIDGDELLA (belonging to xq-dSE3-ubiquitin).*

## S8 Bacterial expression of LBT peptides

The peptides were produced as gene fusions to the GST protein.[12] They were constructed in the vector pGEX-2T, and fusion proteins were expressed in *E. coli* according to the methods described in the GST gene fusion system handbook from GE Healthcare[13]. The original sequence for the **LBT** peptide (YIDTNNDGWYEGDELLA) was obtained from Nitz *et al.*, 2003[14] and the double assymmetric LBT, from now on **daLBT**, (YIDTDNDGWYEGDELYIDTNNDGWYEGDELLA) was designed so as the two **LBT** units were distinct and offer different coordination environments, thus different magnetic energy levels, to make them distinguishable by pulsed EPR.

To generate the GST-TEV-**LBT** and GST-TEV-**daLBT** fusion proteins, the DNA fragment encoding a TEV (Tobacco Etch Virus protease) cut site and the 17-aminoacid **LBT** (or 32-aminoacid **daLBT**) along with BamHI and EcoRI recognition sites at 5' and 3' respectively was ordered as a long oligonucleotide from IDT (Table S8). The TEV cut site (ENLYFQ aminoacidic sequence) was added to facilitate the removal of the GST moiety, which could be required for posterior purification steps. The Forward/Reverse pair of oligonucleotides were annealed as described previously.[15] Briefly, duplex annnealing for both constructions (**LBT** and **daLBT**) was achieved by preparing a mixture of 25 μM final concentration of both primers (F/R), heating at 95ºC for 5 min and letting it cool down slowly to room temperature for 45 min, left on ice for 2 minutes, and finally stored at -20ºC. The annealed duplexes (F/R-BamHI-TEV-LBT-EcoRI and F/R-TEV-daLBT-EcoRI) and vector pGEX-2T were doubly digested with BamHI and EcoRI following ThermoFisher Scientific recommendations. The DNA fragments containing **LBT** or **daLBT** sequences were then ligated into the BamHI-EcoRI pGEX-2T gel-purified digested plasmid, and the product was transformed into XL10-GOLD ultracompetent cells for storage.

Colonies were checked by Sanger sequencing with pGEX-5'-SP and pGEX-3'-SP primers. Plasmids from the verified constructions (both bearing **LBT** or **daLBT** sequences) were recovered and transformed into the *E.coli* expression strain BL21 DE3 pLys. The expression of the peptides was monitored as described in the GST gene fusion system handbook (GE Healthcare)[13] and can be seen in Figure S8.

*Table S8.* *Oligonucleotides used for cloning and verifying sequences.*

| Name | Sequence (5' to 3') |
|---|---|
| F-BamHI-TEV-LBT-EcoRI | ACTGGGATCCGAAAACCTGTATTTTCAGGGCTATATTGATACCAACAACGATGGCTGGTATGAAGGCGATGAACTGCTGGCGTAAGAATTCACTGC |
| R-BamHI-TEV-LBT-EcoRI | GCAGTGAATTCTTACGCCAGCAGTTCATCGCCTTCATACCAGCCATCGTTGTTGGTATCAATATAGCCCTGAAAATACAGGTTTTCGGATCCCAGT |
| F-BamHI-TEV-daLBT-EcoRI | ACTGGGATCCGAAAACCTGTATTTTCAGGGCTATATTGATACCGATAACGATGGCTGGTATGAAGGCGATGAACTGTATATTGATACCAACAACGATGGCTGGTATGAAGGCGATGAACTGCTGGCGTAAGAATTCACTGC |
| R-BamHI-TEV-daLBT-EcoRI | GCAGTGAATTCTTACGCCAGCAGTTCATCGCCTTCATACCAGCCATCGTTGTTGGTATCAATATACAGTTCATCGCCTTCATACCAGCCATCGTTATCGGTATCAATATAGCCCTGAAAATACAGGTTTTCGGATCCCAGT |
| pGEX-5'-SP | GGGCTGGCAAGCCACGTTTGGTG |
| pGEX-3'-SP | CCGGGAGCTGCATGTGTCAGAGG |

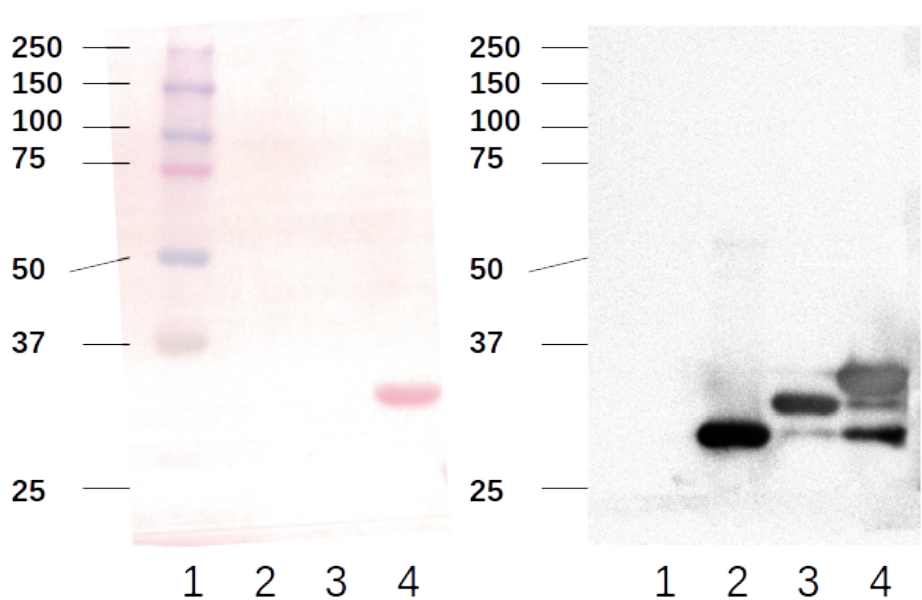

*Figure S8.* Western blot of the GST-LBT and GST-daLBT fusion proteins expressed. Left panel shows a Ponceau S staining of the PVDF membrane, and right panel shows a western-blot developed using anti-GST Horseradish Peroxidase conjugated antibody (GE Healthcare) with SuperSignal West Pico Chemilluminescent substrate (Thermo Fisher) as recommended by manufacturers. Molecular weight markers are shown on the left. Lanes are as follow: 1) molecular weight marker (BioRad Precision Plus Protein Kaleidoscope Standards), 2) Elution of GST, 3) Elution of GST-TEV-LBT, 4) Elution of GST-TEV-daLBT.

# S9 Design of a nonanuclear biomolecular spin qubit

## S9.1 Context and motivation

Let us assume a multi-qubit biomolecule, where the different spin qubits communicate with each other via dipolar coupling. In that situation, it is obvious that different relative orientation of the different spins and their easy axes of magnetization will result in different effective spin Hamiltonians. A given qubit pair, for example, can either be coupled ferro- or antiferromagnetically, or not at all; and in fact if one is able to choose the position and orientation of n spins, the ground state can be chosen among a large variety of degeneracies. It is equally clear that, for any desired quantum operation, some of these spin structures will be better suited than others.

Thus, this general strategy allows for many variations of short chains where members interact by dipolar interactions, with the best configurations depending on the desired quantum algorithm. The key advantage of this proposal for the quantum theoretician is the fact that by choosing the position and orientation of a number of LBTs, it is in principle posible to design and build the hardware that has the desired connectivity between qubits (and thus, within some limitations, the desired spin Hamiltonian), something that cannot currently be achieved with other approaches.

Of course, currently it is not obvious how to situate an arbitrary number of spins in these peptide frameworks (for more on these see the next section of Supporting Information, S11). What we intend to do here is merely to demonstrate that it is possible to link a rather long chain of two different kinds oflanthanide coordination environments, disregarding for the moment the key problem of their positions and orientations, which is related to a wise design of the linkers.

## S9.2 Experimental approach

In a similar manner as the **daLBT** peptide, an aminoacidic sequence bearing two different units of LBT termed 0 and 1 respectively (the corresponding DNA and protein sequence can be seen in table S9.1) was designed so as to have a total of nine units. Using this notation, the chimeric peptide **daLBT** then would be called 's10' (standing for 'sequence 10'). The proposed LBT string containing nine units (0 and 1) was designed with the following sequence 010011101 ('s01001110'). This sequence is arbitrary and was chosen to illustrate the fact that there is no symmetry (or homogeneity) restriction, as it would often happen in coordination chemistry. In particular, if grouped in LBT triplets and read in binary, 010, 011, 101 stands for 2, 3, 5, the first three prime numbers. Each LBT was placed between short linkers (in this context the linkers used are sequences 5 or 7 aminoacids long) composed of small, non-charged aminoacids, which are designed to allow the LBT units to fold without unwanted interactions among them. The different linkers suggested are listed in the table S9.2. We aimed for this in order to obtain a high affinity binding of lanthanide cations to our designed protein.

| LBT unit | | Sequence (5' to 3') (N-t to C-t) |
|---|---|---|
| 0 | DNA | TATATTGATACCAACAACGATGGCTGGTATGAAGGCGATGAACTGCTGGCG |
| | protein | Y I D T N N D G W Y E G D E L L A |
| 1 | DNA | TATATTGATACCGATAACGATGGCTGGTATGAAGGCGATGAACTG |
| | protein | Y I D T D N D G W Y E G D E L |

**Table S9.1.** *LBT units used for designing a lanthanide-bearing nonanuclear spin qubit.*

One could use standard tecniques of recombinant protein expression for the preparation of this peptide with nine LBT units. Also it should be easy to prepare polypeptides containing 3 to 9 LBT units due to the the use of unique linkers for the spacing of the units. The unique spacers allow for the specific PCR amplification of DNA encompassing 1, 2, 3, 4, 5, 6, 7, 8 or 9 LBT units, which in turn could be sub-cloned and purified in *E. coli*. The linkers were all designed with a GASAG amino acid sequenced (unique DNA sequences) to facilitate independent folding of each LBT.

In short, the DNA sequence coding this protein of interest (see box S9) would be placed under the control of an inducible Lac promoter (expression only occurs when Isopropyl β-D-1-thiogalactopyranoside (IPTG) is present in the culture media)[16] and inserted into an *E. coli* commercial plasmid containing an Ampicillin resistance marker. The cloning procedure would be performed in a similar way as presented for the dimer in section S7. Then, the plasmid-containing bacteria could be grown in the presence of IPTG, the 's010011101' fusion protein expressed, and purified from the culture. The purificaction is usually enabled by peptidic tags. These tags can either be fused in N- or C-terminal position to the protein of interest in the context of the plasmid, allowing to perform Affinity Chromatography to purify the cloned protein.

*Table S9.2. Linkers employed for the designed of the nonanuclear LBT peptide.*

| # Linker | | Sequence (5' to 3') (C-t to N-t) |
|---|---|---|
| 1 | DNA | ggcagcggcgcgagcgcgggc |
|   | protein | G S G A S A G |
| 2 | DNA | gcgggcagcggcgcg |
|   | protein | A G S G A |
| 3 | DNA | ggcggcagcgcggcg |
|   | protein | G G S A A |
| 4 | DNA | gcggcgggcgcggcg |
|   | protein | A A G A A |
| 5 | DNA | ggcggcgcgggcggc |
|   | protein | G G A G G |
| 6 | DNA | gcggcgagcggcggc |
|   | protein | A A S G G |
| 7 | DNA | gcgagcggcagcgcg |
|   | protein | A S G S A |
| 8 | DNA | ggcagcgcgagcggc |
|   | protein | G S A S G |
| 9 | DNA | ggcgcgggcgcgagc |
|   | protein | G A G A S |
| 10 | DNA | taatagtgaaattgg |
|   | protein | stop codons |

```
ggcagcggcgcgagcgcgggcTATATTGATACCAACAACGATGGCTGGTATGAAGGCGATGAACTG
CTGGCGgcgggcagcggcgcgTATATTGATACCGATAACGATGGCTGGTATGAAGGCGATGAACTG
gcggcgggcgcggcgTATATTGATACCAACAACGATGGCTGGTATGAAGGCGATGAACTGCTGGCG
gcggcgggcgcggcgTATATTGATACCAACAACGATGGCTGGTATGAAGGCGATGAACTGCTGGCG
ggcggcgcgggcggcTATATTGATACCGATAACGATGGCTGGTATGAAGGCGATGAACTGgcggcg
agcggcggcTATATTGATACCGATAACGATGGCTGGTATGAAGGCGATGAACTGgcgagcggcagc
gcgTATATTGATACCGATAACGATGGCTGGTATGAAGGCGATGAACTGggcagcgcgagcggcTAT
ATTGATACCAACAACGATGGCTGGTATGAAGGCGATGAACTGCTGGCGggcgcgggcgcgagcTAT
ATTGATACCGATAACGATGGCTGGTATGAAGGCGATGAACTGtaatagtgaaattgg
```

**Box S9.** *DNA sequence (5' to 3') corresponding to the proposed lanthanide-coordinating nonanuclear LBT protein.*

# S10 Further approaches for organising Rare Earths employing biomolecules

A long-term requirement for functional quantum devices is to produce highly complex quantum structures with a resolution scale below the nanometer and total sizes above than the micrometer. Indeed, the fabrication of scalable quantum devices will depend on the capability for the dense, programmable, atomic-resolution organisation of different kinds of active centres with a nanometric resolution, to allow for controlled interactions between them. Indeed, organisation predates function. A unique feature of biomolecules is the possibility of self-organization into highly complex structures. This points toward the possibility of using biomolecules for the organization of functional quantum building blocks.

As the structure of DNA can be controlled up to the level of folding it like origami,[17] attaching Rare Earth (RE) ions to DNA would be the obvious option. Back in the 80s the interaction of the RE ions with nucleic acids was determined, including DNA and RNA, and it was found to be not selective enough for our purposes.[18] Also there were studies of the direct interaction of RE ions with histone proteins,[19] which again show limited specifity. This is why we focus on Lanthanide Binding Tags (LBTs) which are more specific.

For the ordering of LBTs at least three approaches are conceivable:

(1) using protein dimers with specific interactions to bind LBT pairs (or, in general tuples)

(2) ordering on nucleosome proteins + DNA, and

(3) ordering on proteins with repetitive sequences.

In the first approach, one would take advantage of an already described protein dimer to promote the interaction of two distinct lanthanide binding tags. Each peptidic sequence would then be genetically inserted into a monomer of the previously mentioned dimer. The resultant chimeric proteins would then be expressed in *E. coli* for their purification prior to the generation of the required crystals. In this approach it is crucial the insertion of the **LBT** in a position that allows the close interaction with its partner, an imperative for dipolar communication between the two spin qubits. To ensure the proper location of the tag, a dimer whose 3D atomic structure has already been solved by X-ray crystallography should be selected. At the present time 6205 dimer structures have been resolved, which give a large pool to choose from. Note that the structure of the chimeric protein must be solved to ensure that the LBT labeling has not disrupted the overall protein structure. This approach is extensible to trimers and tetramers with multiple combinations
of Lanthanide Binding Tags.

The nucleosome, which is key for the second approach, is the fundamental repeating element of chromatin (nuclear structure that contains the genetic material in the eucharyotic cell). It comprises between 157 and 240 base pairs (bp) of DNA, the four core histone proteins (H2A, H2B, H3 and H4) and the linker histone H1. The nucleosome core contains 147 bp of DNA supercoiled in 1.67 left-handed turns around a core histone octamer. The octamer consists of one tetramer of $(H3-H4)_2$, and two dimers of (H2A-H2B), and has a molecular weight of 110 KDa. The possiblility of using the biomimetic self-organization of proteins that stably bind to DNA to form the nucleosome core superstructure, is something one can benefit from. Histones are small (11-16 KDa), basic proteins, with two differentiated domains: a flexible N-terminal tail with basic properties 15-30 aminoacids long, and a C-terminal globular domain (approx. 65 aminoacids) that contains a structural motive called histone-fold, important for molecular contacts within the histone octamer. The tagging of the core histones with LBTs is feasible using standard molecular biology tools.

Nucleosome assembly in vitro has been the object of many studies, so reliable protocols are readily available in the literature.[20] Core histone octamers that are repetitively spaced along a DNA molecule are known as nucleosomal arrays. Nucleosomal arrays have been reconstituted from recombinant histone octameres, using as template DNA well known nucleosome positioning sequences (NPS).[21] These DNA sequences are able to strongly position a nucleosome in vitro, suggesting the possibility that favorable instrinsic signals might reproducibly structure chromatin fragments which allows for the organization of nanoobjects. These NPS are vital for this approach as they provide specific docking sites for the histone octamer (our nanoobjects) on in vitro reconstitution assays (also called chromatin assembly assays). Nucleosome assembly can be achieved then, by combining core histone octamers that will bind to tandemly repeated nucleosome positioning sequences. Widom 601 sequence, 5s rDNA or MMTV-B sequence are NPS of about 170 bps that are currently used in nucleosome assembly in vitro.[22] The reconstitution assays allow the assembly of a compositionally uniform and precisely positioned nucleosomal array that could act as the scaffold of our qubit-LBT bearing histones. Histone binding in nucleosome reconstitution is not expected to be strongly affected by the addition of an LBT in its N-terminal end (or other positions within the protein), as this tag is only 15 aminoacids long.

The third approach would be based on proteic tandem repeats as LBT scaffold. Up to 14% of all proteins contain repetitive regions.[23] Repeats vary considerably in size, order and complexity. In some cases, protein repeats can occur tandemly in sequence forming integrated assemblies units when viewed as three-dimensional structures (for example TALEN). Such repeats are essentially defined by their multiplicity and thus differ from both domains and motifs since these can occur in a variable number of repeats without altering the integrity, activity or stability of the overall protein.[24] The structure of many of these proteins has already been solved, giving an atomic (Å) resolution of the location of the repeats and the distance between them. Thus, these proteins represent an splendid scaffold for the insertion of Lanthanide Binding Tags (LBT) in an ordered manner.

Knowing the tridimensional structure of the protein once it forms a crystal allow us to design LBTs in a specific location and calculate the distance with other LBT within the same molecule. Therefore for the success of this approach it is imperative to confirm that the insertion of LBTs does not alter the crystal structure of the scaffold protein. Using proteins whose structure has already been solved can facilitate and speed up the work tremendously, since the protocols to purify and crystallize the proteins have already been described.

Finally, one should consider the possibility of combining the organising power of proteins with the advantages of polyoxometalates for the design and stabilization of molecular spin qubits. These inorganic molecules provide a rigid structure for lanthanide ions, with very few nuclear spins, and which have been proven to result in promising molecular spin qubits.[25–27] Interestingly, different polyoxometalates, including those of Lindqvist-, Keggin- and Wells–Dawson-types, have been found to bind regioselectively to positively-charged regions of different proteins.[28] So far, this research has been oriented toward the regioselective hydrolysis of proteins, but it is also a potential pathway towards the biomolecular organisation of polyoxometalate spin qubits, provided that autonomously folding, positively charged regions can be found. These could then be structured employing the same strategies as described above for LBTs.

# S11 Preparation of LBTC Self-Assembled Monolayers (SAMs)

The stability of the polypeptide relies in the pH of the media, which implies the use of aqueous buffered solutions where reducing agents as 2-mercaptoethanol are commonly added to avoid the oxidation of the thiol groups. In order to be sure that the **LBTC** keeps its integrity during the assembling processes we decided to use the standard solution with HEPES (4-(2-hydroxyethyl)-1-piperazineethanesulfonic acid) as buffer, but we were forced to substitute the 2-mercaptoethanol by a new reducing agent without thiol functional groups. The thiol group of the reductant competes with the cysteine thiol group, which is expected to be bonded to the Au atoms in the assembling process on the metallic surface, driving the formation of a self-assembled monolayer (SAM). To avoid this competition problem 2-mercaptoethanol was substituted by tris(2-carboxyethyl) phosphine (TCEP). This phosphine can keep the cysteine from forming di-sulphide bonds (oxidation problems), but it has not a special affinity for the Au surface.

However, the use of TCEP can mean a different problem. When a metallic cation like $Tb^{3+}$ is added to the solution in order to form **TbLBTC** complex, the possibility of a chelating effect of the TCEP (three carboxylic groups), cannot be ruled out. To guarantee that **LBTC** coordinates the $Ln^{3+}$ cation when **LBTC**, TCEP and $Ln^{3+}$ ions are in the same solution, we worked always with and excess of the lanthanide and we checked by UV spectroscopy that the expected luminescence emission at 544 nm induced by the **LnLBTC** complex when it is excited at 280 nm, was present (Figure S1.b).

In order to form a SAM, a simple immersion method was used. This is the standard approach for the formation of alkanethiol SAMs on Au, and only small modifications were introduced. First, the gold substrate was cleaned and activated by piranha treatment, then the substrate was incubated in a buffered **TbLBTC** aqueous solution (HEPES 10mM pH 7.0, NaCl 100mM, $TbCl_3$ 0.11 mM, TCEP 0.5mM, **LBTC** 0.1mM) overnight, and finally in order to remove non-chemisorbed material that could remain on the surface, the sample was introduced in water for two hours and copiously rinsed.

After the treatment, the topology of the functionalized substrate was studied by Atomic Force Microscopy (AFM). This analysis is necessary in order to be sure that there are not aggregates covering the surface. Their presence would mask the compositional information obtained by any surface technique. After functionalization, the roughness remains almost unchanged suggesting a homogeneous coverage, as can be seen in Figure 4.a in the main text, where the topographic AFM images of a substrate functionalized with a **TbLBTC** SAM is compared with a reference sample (incubated overnight in a buffered solution without LBTC).

To determine the presence of **TbLBTC** on the gold surface Matrix-assisted laser desorption/ionization time of flight (MALDI-TOF) spectrometry and X-ray photoelectron (XPS spectroscopy) were performed (see sections S12, and S13).

# S12 Characterization of the LBTC-SAMs by Matrix-Assisted Laser Desorption/Ionization-Time of flight (MALDI-TOF)

MALDI is a soft ionization technique commonly used for the mass spectrometry analysis of bio- and large organic molecules. The spectrum shows charged fragments of the macromolecule that are obtained in the gas phase after ionizing the sample. Soft interactions like electrostatic ones between LBTC and Ln are easily broken during this ionization process, and only LBTC fragments (charge +1) are expected in the mass spectra. As shown in Figure S12 the main fragment of the SAM appears at 1629, there are also two clear peaks at 1907 and 1929 m/z. The peak at 1629 belongs to the fragment DTNNDGWYEGDELC and the peaks at 1907 and 1929 m/z correspond to the molecular weight of the LBTC and the LBTC with an additional $Na^+$ respectively. This result is in good agreement with a reference spectra performed on solution, where the peak of the adduct LBTC-Na is visible. This proved that the integrity of the molecule is preserved after its attachment on the Au surface.

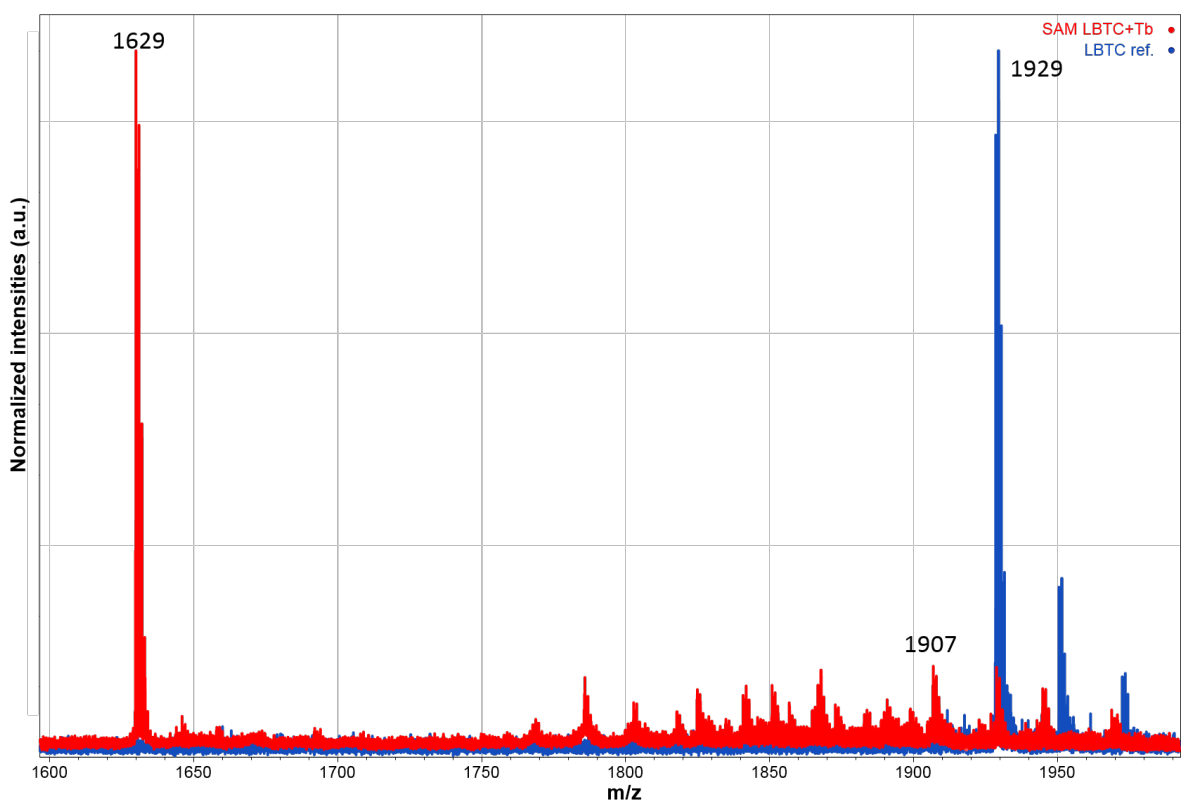

***Figure S12.*** *MALDI-TOF spectrum of **LBTC** in solution (blue) and **TbLBTC** grown on top of a gold surface (red).*

# S13 Characterization of the LBTC-SAMs by X-ray photoemission spectroscopy (XPS)

Due to the impossibility to identify the **TbLBTC** complex by mass spectrometry, X-ray photoemission spectroscopy (XPS) was used. XPS spectra clearly show the presence of Tb3d peaks (1277 and 1242 eV) denoting the presence of the lanthanide on the surface. The presence of the binding tag is also probed by S2p (162 eV, thiolate bonded to Au) and N1s (400 eV, amino groups) peaks (Figure S13).

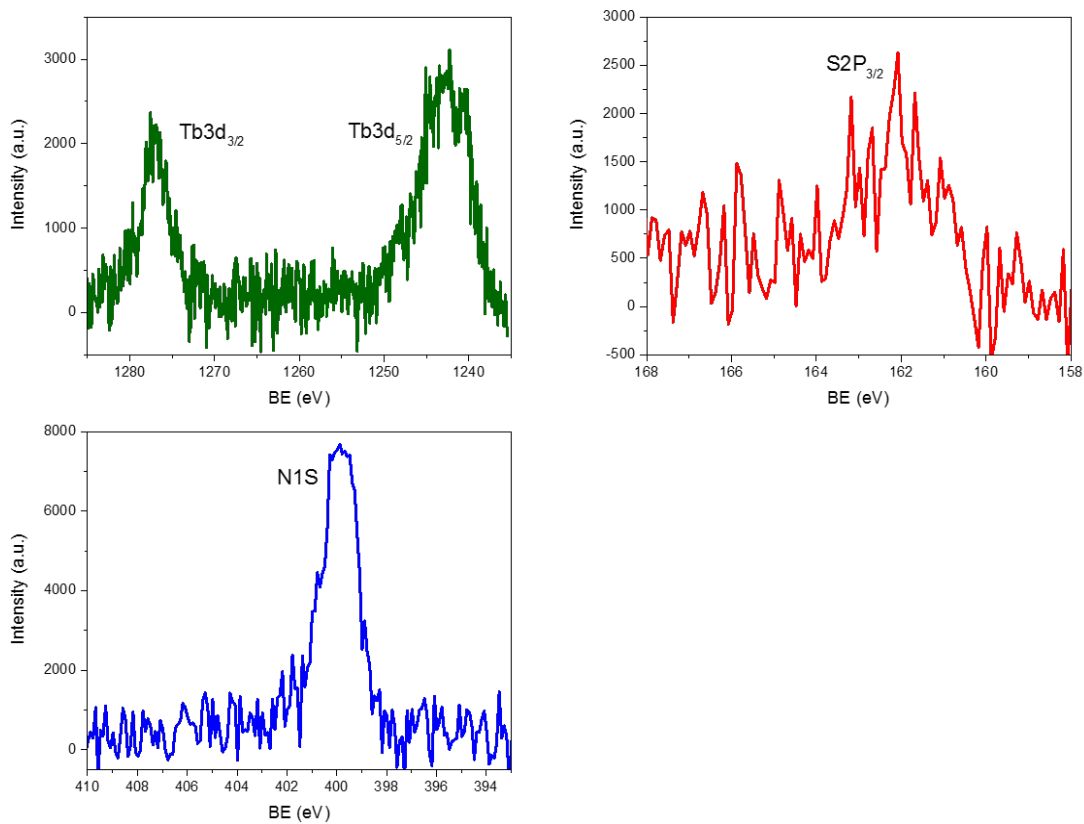

*Figure S13. XPS spectrum of Tb (green), S (red) and N (blue) of a **TbLBTC** monolayer.*

# S14 Coverage evaluation by means of Quartz Crystal Microbalance (QCM) Experiments

In this work, for a typical experiment, a freshly cleaned Au covered piezoelectric quartz crystal was immersed in a liquid cell where a flow of different liquids at controlled rates was pumped while the oscillation frequency of the crystal was continuously monitored. Several cycles of pure water and buffered aqueous solution were performed until the frequency was stabilized. Then right after a buffered solution step, **TbLBTC** or **LBTC** solution was injected during ca. 60 min to ensure a constant frequency value. Finally, a flow of the buffered aqueous solution was introduced to remove physisorbed **TbLBTC** or **LBTC** on the crystal.

$$\Delta f = -\frac{2f_0^2}{A\sqrt{\rho\mu}} \Delta m$$

Sauerbrey's equation was used to estimate the relation between resonant frequency and mass loading of electrodes.[29]

where $f_0$ is the resonant frequency of QCM, $\Delta f$ is the change in the resonant frequency of quartz microbalance due to the mass change ($\Delta m$) of electrodes. $A$ is the active area, $\rho$ is the density and $\mu$ is the shear modulus of the quartz crystal provided by the manufacturer.

To calculate the theoretical coverage of the surface two different models have been used that give rise to a range of coverage values:

-Model A: the molecule is considered as a cube of 1.7 nm side

-Model B: the molecule is considered a sphere which projection on the surface is a circle of 1.6nm diameter, with a hexagonal compact arrangement, giving rise to a coverage density of $\pi/2\sqrt{3}$ (ca. 0.9069).

It is important to clarify that the molecular weights of the **LBTC** with or without $Tb^{3+}$ are not precise as the counter ions present and hydratation molecules are unknown. We have considered the simplest situation where protons compensate negative charges of the polypeptide, this means that the final coverages estimated are the highest values expected.

The coverage calculated from **LnLBTC** experiments are described in the main text giving rise to a range of coverage between 80-116%. In spite of the absence of lanthanide in the **LBTC**, its tertiary structure keeps its round shape (based on **LBTC** crystal structure), and the same theoretical shape can be estimated. As observed in Figure S14, when **LBTC** is anchored on the electrode, the frequency shift measured is one third higher than for **TbLBTC** (30Hz vs 20Hz). This large value indicates a coverage between 145-189%. This could indicate that with the Tb complex the globular-like structure remains unchanged when the thiol groups are attached to the gold and almost all the surface is covered. However, the large amount of material estimated when the $Tb^{3+}$ is not present in the solution only can be justify by two different explanations: (i) intermolecular polypeptide-polypeptide interactions are stronger than intramolecular interactions forcing a less globular shape and permitting a higher coverage of the surface, or (ii) a second layer is formed on top of the first layer. From our point of view, the second option is less probable as there is no reason to think that a second layer can be formed with **LBTC** but it cannot with **TbLBTC**, so it should be observed in both cases.

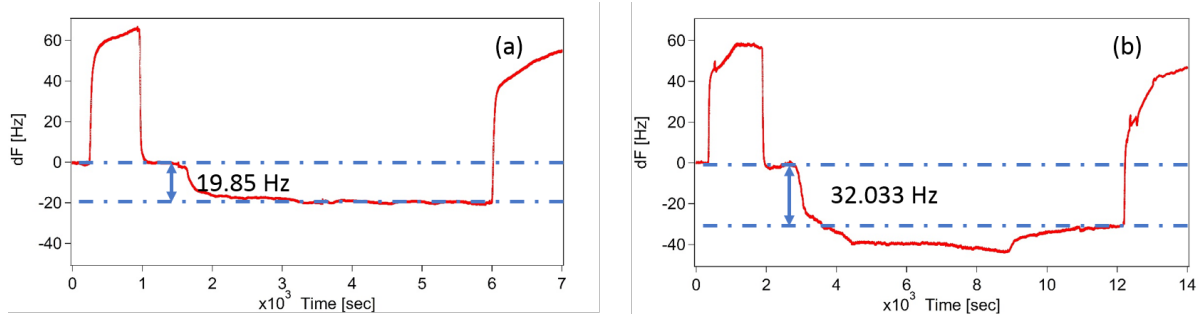

*Figure S14.* QCM measures of **LBTC** coordinating Tb (a) and **LBTC** alone (b).

# S15 Transport Calculations on LBTC between two gold electrodes

First-principles calculations are performed using the SMEAGOL code[30,31] that interfaces the non-equilibrium Green's function (NEGF) approach to electron transport with the density functional theory (DFT) package SIESTA.[32]

In our simulations the transport junction is constructed by placing the polypeptide between two Au (111)-oriented surfaces with 7x7 cross section. This mimics a standard transport break-junction experiment with the most used gold surface orientation. The choice of a gold electrode arises from the stability of the sulfur-gold bond that ensures the best attachment between the cysteine and the surface. The initial S-surface distance was set to 2.0 Å with the S atom located at the 'hollow site', the most stable absorption position discussed in literature.[33] Thus, the entire structure is then relaxed until the maximum atomic forces are less than 0.01 eV/Å. A real space grid with an equivalent plane wave cutoff of 200 Ry (enough to assure convergence) has been used to calculate the various matrix elements. During the calculation the total system is divided in three parts: a left-hand side lead, a central scattering region (SR) and a right-hand side lead. The scattering region contains the molecule as well as 4 atomic layers of each lead, which are necessary to relax the electrostatic potential to the bulk level of Au.

The convergence of the electronic structure of the leads is achieved with 2x2x128 Monkhorst-Pack k-point mesh, while for the SR one sets open boundary conditions in the transport direction and periodic ones along the transverse plane, for which an identical k-point mesh is used (2x2x1 k-points). The exchange-correlation potential is described by the GGA (PBE) functional. The Au-valence electrons are represented over a numerical s-only single-ζ basis set that has been previously demonstrated to offer a good description of the energy region around the Fermi level.[34] In contrast, for the other atoms (S, C, V and O) we use a full-valence double-ζ basis set. Norm-conserving Troullier-Martins pseudopotentials[35] are employed to describe the core-electrons in all the cases.

Finally, the spin-dependent current, $I_\rho$, flowing through the junction is calculated from the Landauer-Büttiker formula,[36]

$$I_\sigma(V) = \frac{e}{h} \int_{-\infty}^{+\infty} T_\sigma(E,V)[f_L(E - \mu_L) - f_R(E - \mu_R)]dE$$

where the total current $I_{tot}$ is the sum of both the spin-polarized components, $I_\sigma$, σ = spin up /spin down. Here $T_\sigma(E,V)$ is the transmission coefficienti and $f_{L/R}$ are the Fermi functions associated to the two electrodes chemical potentials, $\mu_{L/R} = \mu_o \pm V/2$, where $\mu_o$ is the electrodes common Fermi level.

## S16 Magnetic Field Calculations in Space Points

Here, we explain how to calculate the magnetic field induced in a space point (in this case, each atom of the LBT chain) by a magnetic ion displaying a magnetic moment. Let $\vec{r}$ be the relative position of the space point respect to the position of the magnetic ion, and $\vec{m}$ be te magnetic moment associated to the magnetic ion. The expressions of these vectors are:

$$\vec{r} = (r_x, r_y, r_z); \vec{m} = \mu_B \left( g_x \langle \hat{J}_x \rangle, g_y \langle \hat{J}_y \rangle, g_z \langle \hat{J}_z \rangle \right)$$

where $r_\alpha$ are cartesian coordinates, $g_\alpha$ are the Landé tensor main components of the **LnLBT** complex and $\langle \hat{J}_\alpha \rangle$ are the x, y, z component expectation values of the spin-orbit $\hat{J} = \hat{L} + \hat{S}$ operator when the magnetic ion is in one of the *(2J + 1) (2I + 1)* eigenstates, being *I* the nuclear spin of the ion. We chose the ground eigenstate, and the corresponding expectation values are calculated using the SIMPRE1.2.[4] Then, the magnetic field induced at the position by the magnetic moment is:

$$\vec{B}(\vec{r}) = \frac{\mu_0}{4\pi} \frac{1}{|\vec{r}|^5} \left( 3(\vec{m} \cdot \vec{r})\vec{r} - |\vec{r}|^2 \vec{m} \right)$$

After some algebra, one finds that each component of $\vec{B}(\vec{r})$ can be written as:

$$\left( \vec{B}(\vec{r}) \right)_\alpha = \frac{\mu_0 \mu_B}{4\pi} \frac{1}{|\vec{r}|^5} \left( 3 r_\alpha S - |\vec{r}|^2 g_\alpha \langle \hat{J}_\alpha \rangle \right); \alpha = x, y, z$$

and the module of this induced magnetic field is:

$$\left\| \vec{B}(\vec{r}) \right\| = \frac{\mu_0 \mu_B}{4\pi} \frac{1}{|\vec{r}|^4} \left( 3S^2 + S' |\vec{r}|^2 \right)^{1/2}$$

where *S* and *S'* are:

$$S := \sum_{\alpha=x,y,z} g_\alpha \langle \hat{J}_\alpha \rangle r_\alpha; S' := \sum_{\alpha=x,y,z} \left( g_\alpha \langle \hat{J}_\alpha \rangle \right)^2$$

The calculations were performed selecting a static external magnetic field of 0.1 T and 1.0 T to calculate the expectation values $\hat{J}_\alpha$.

## S17 Crystallization attempts of LBT peptides

Crystallography of LBT proved challenging. Peptide crystallization was pursued by the vapor diffusion method using the sitting drop technique by mixing 0.4 μl of peptide (2.5 and 5 mg/ml in 10 mM HEPES and 1.2 mM $TbCl_3$ pH 7.0) with 0.4 μl of different reservoir solutions in 96-well MRC-type crystallization plates. Reservoir solutions were based on previous reported crystallization conditions (3.7 mM NaCl, 100 mM Tris at pH 6.5 and 40% t-butanol) as well as in different sparse matrix screens.[37] Sparse matrix screens used for peptide crystallization were JBScreen Classic HTS I and II, JBScreen JCSG ++ and JBScreen PACT ++ from Jena Bioscience (Germany). We tested all resulting crystals at the synchrotrons Alba (Barcelona, Spain) and Diamond Light Source (Didcot, UK), but in every case they corresponded to salts. Thus, we were unable to reproduce the previous reported crystals obtained by the microbatch technique.

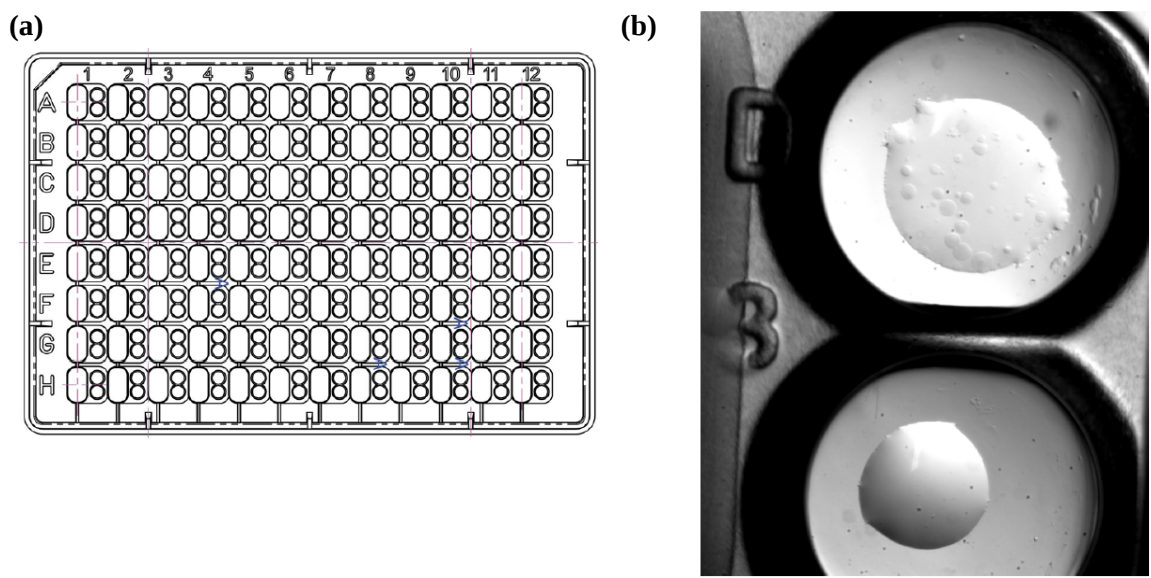

*Figure S17.* (a) Scheme of the typical MRC-type plate we used for crystallization assays. (b) View of a pair of wells with spherulites in one of them from one of our assays.